\documentclass[prb,showpacs,byrevtex,amsmath,amssymb,twocolumn]{revtex4}
\usepackage{graphicx}% Include figure files
\usepackage{dcolumn}% Align table columns on decimal point
\usepackage{bm}% bold math
\begin{document}
%\draft
\preprint{PREPRINT}

\title[Short Title]{Image charges in spherical geometry: Application to colloidal systems}
% Force line breaks with \\

%1st Author
\author{Ren\'e Messina}
%\thanks{Also at Physics Department, XYZ University.}%Lines break automatically or can be forced with \\
\email{messina@mpip-mainz.mpg.de}% REVTEX 4

\affiliation{Max-Planck-Institut f\"ur Polymerforschung, Ackermannweg 10,
  55128 Mainz, Germany}

\date{\today}% It is always \today, today, but you may specify any date with
             % \date.

\begin{abstract}
The effects of image charges (i.e., induced surface charges of polarization) in
spherical geometry and their implication for charged colloidal systems
are investigated.
We study analytically and exactly a single microion interacting
with a dielectric sphere and discuss the similarities and discrepancies
with the case of a planar interface.
By means of extensive Monte Carlo (MC) simulations, we study within the framework of the
primitive model the effects of image
charges on the structure of the electrical double layer. Salt-free environment as
well as salty solutions are considered.
A remarkable finding of this study is that
the position of the maximum in the counterion density
(appearing at moderately surface charge density)  remains quasi-identical,
regardless of the counterion valence and the salt content,
to that obtained within the \textit{single}-counterion system.
\end{abstract}

% PACS, the Physics and Astronomy Classification Scheme.
\maketitle

%
%HWG with the MAIN TEXT
%

%%%%%%%%%%%%%%%%%%%%%%%
\section{Introduction}
%%%%%%%%%%%%%%%%%%%%%%%

In charged colloidal systems electrostatic effects, and especially
the structure of the electrical double layer,  often play a crucial role
in determining their physico-chemical properties. It is well known
that charged colloids (i.e., macroions) have typically a low dielectric
constant ($\varepsilon_r\approx2-5$) which is much smaller than that of
the surrounding solvent (e.g., for water $\varepsilon_r\approx80$).
In most of the theoretical works, this dielectric discontinuity is ignored.

Nevertheless, a few studies  have addressed the effects of image charges
(i.e., image forces stemming from the dielectric discontinuity) on
the counterion distribution for  planar geometry which is closely related to
our problem. An electrolyte close to a charged wall
\cite{Torrie_JCP_1982,Torrie_JCP_1984}  or confined between two charged plates
\cite{Bratko_CPL_86} had been the subject of MC simulations.
Similar systems have also been investigated by integral-equation
\cite{Croxton_CanJC_81,Kjellander_CPL_84,Kjellander_JCP_85}
and mean field theories.
\cite{Outhwaite_JCSFT_83,Netz_PRE_1999,Gruenberg_JPhysCondM_2000}

As far as the spherical geometry is concerned, much less literature is available.
Counterion distributions with image forces in salt-free environment had been
investigated by MC simulations. \cite{Linse_JPC_1986}
There an oversimplified approximation for the
treatment of the image forces was used.
The main conclusions however
remain correct on a very qualitative level of description.

The aim of this paper is to provide a detailed analysis of
the image forces in spherical geometry and their effects on the
structure of the electrical double layer.
The remainder of this article is set out as follows.
Section \ref{Sec.Theory} corresponds to the analytical part of the paper.
We first briefly present the general theoretical
background of the concept of image charges in spherical geometry.
Then we  apply it to colloidal systems to compute (exactly) some
relevant observables and discuss our results.
Section \ref{Sec.MC} is devoted to the computational details of
our MC simulations.
In Sec. \ref{Sec.SIMU-results} we present our simulation results for
salt-free environment as well as salty solutions where image forces are
explicitly taken into account with no  approximation.
Finally, Sec. \ref{Sec.conclu} contains brief concluding remarks.

%%%%%%%%%%%%%%%%%%%%%%%%%%%%%%%%%%%
\section{Theory\label{Sec.Theory}}
%%%%%%%%%%%%%%%%%%%%%%%%%%%%%%%%%%%

In this part we  mainly study the interaction of a \textit{single
excess} charge with a  dielectric sphere.
We briefly present the formalism of the dielectric model for spherical interfaces
and discuss some important electrostatic properties. Such a system captures
the underlying physics of image forces in spherical geometry.
Moreover, a systematic quantitative comparison with the planar geometry is undertaken.

\subsection{Poisson equation with azimuthal symmetry }

\begin{figure}
\includegraphics[width = 8.0 cm]{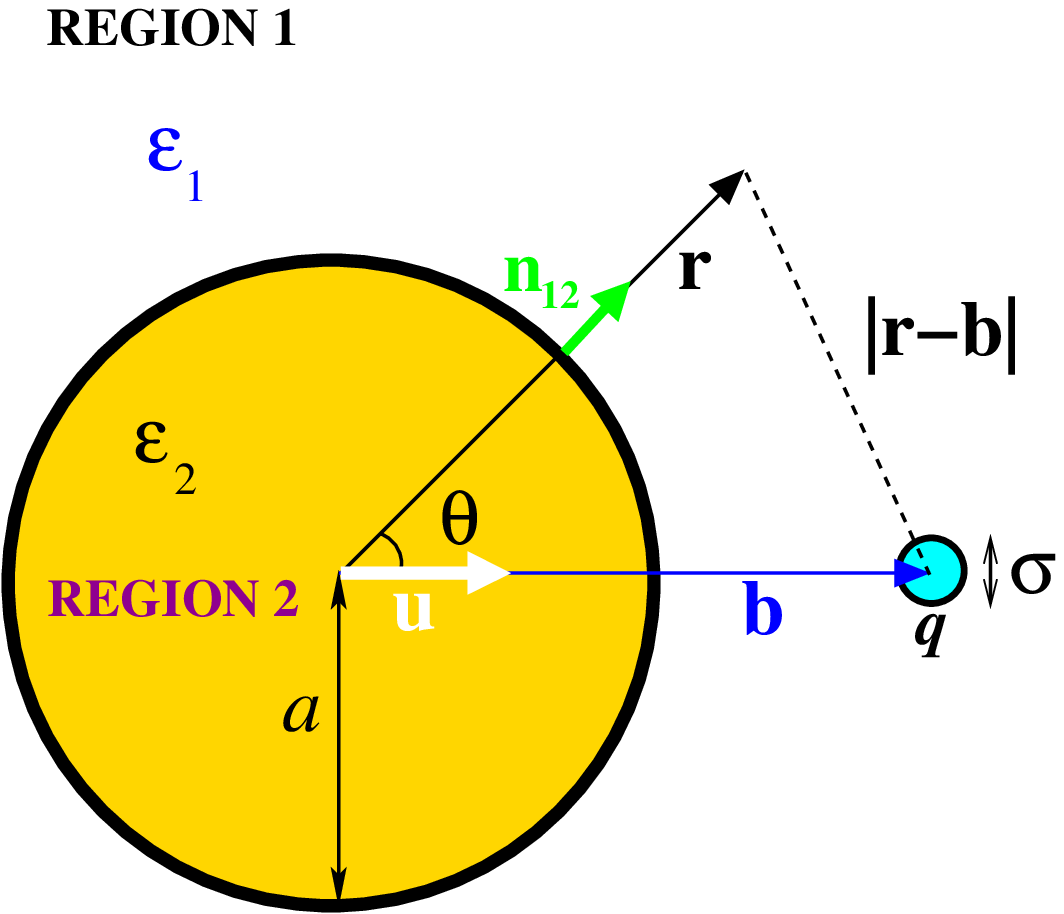}
\caption
{
Model for a dielectric sphere (colloid) of dielectric constant $\varepsilon_2$
embedded in an infinite medium characterized by a different dielectric
constant $\varepsilon _{1}$.
An excess charge ($q$) is located near the boundary outside the
spherical particle.
This is a two-dimensional representation of the
three-dimensional system.
}
\label{fig.setup-point}
\end{figure}

The model system is sketched in Fig. \ref{fig.setup-point}. Consider
an \textit{uncharged} dielectric sphere of radius $a$ and dielectric constant (relative
permittivity) $\varepsilon_2$ embedded in an infinite dielectric
medium (region 1) characterized by $\varepsilon_1$.
A single excess charge of magnitude $q$ is located outside the
dielectric sphere at a distance $b=|\mathbf{b}|$ from its center.

The central problem is to determine the electrostatic potential
$\Phi (\mathbf{r})$ at any point in the space. This is achieved by
solving the Poisson equation which reads

%%%%%%%%%%%%%%%%%%%%%%%
\begin{equation}
\label{Eq.poisson}
\Delta \Phi (\mathbf{r})=-\frac{\rho (\mathbf{r})}{\varepsilon},
\end{equation}
%%%%%%%%%%%%%%%%%%%%%%%
%
where $\rho (\mathbf{r})$ is the volume charge density and
$\varepsilon =\varepsilon_0\varepsilon_i$
with $\varepsilon _{0}$ being the vacuum permittivity and $i=1,2$.
Since here $\rho (\mathbf{r})=q\delta (\mathbf{r}-\mathbf{b})$
and taking into account the azimuthal symmetry, Eq. (\ref{Eq.poisson})
reduces (for $\mathbf{r}\neq \mathbf{b}$) to the Laplace equation

%%%%%%%%%%%%%%%%%%%%%%
\begin{equation}
\label{Eq.laplace}
\Delta \Phi (r,\theta)=
\frac{1}{r^{2}}\frac{\partial }{\partial r}
\left( r^{2}\frac{\partial \Phi}{\partial r}\right) +
\frac{1}{r^{2}}\frac{1}{\sin \theta }\frac{\partial }{\partial \theta }
\left( \sin \theta \frac{\partial \Phi }{\partial \theta }\right) =0,
\end{equation}
%%%%%%%%%%%%%%%%%%%%%%
%
where $\theta$ is the angle between $\mathbf{r}$ and $\mathbf{b}$
(see Fig. \ref{fig.setup-point}) and $r=|\mathbf{r}|$. The general
solution of the Laplace equation with azimuthal symmetry is given
by \cite{Kirkwood_JCP_1934,Spiegel_Fourier_Book_1974,Jackson_book_1975}

%%%%%%%%%%%%%%%%%%%%%%
\begin{equation}
\label{Eq.FI-general}
\Phi (r,\theta )=
\sum ^{\infty }_{l=0}\left[ M_lr^l+N_l\frac{1}{r^{l+1}}\right] P_l(\cos \theta),
\end{equation}
%%%%%%%%%%%%%%%%%%%%%%
%
where $P_{l}(\cos \theta )$ is the associated Legendre polynomial
of order $l$.

Inside the dielectric sphere (region 2) the electrostatic potential
$\Phi_2(\mathbf{r})$  must be finite at $r=0$ so that $N_{l}=0$
in Eq. (\ref{Eq.FI-general}), and hence

%%%%%%%%%%%%%%%%%%%%%%
\begin{equation}
\label{Eq.FI-in}
\Phi_2(r,\theta )=\sum ^{\infty }_{l=0}A_lr^lP_l(\cos \theta).
\end{equation}
%%%%%%%%%%%%%%%%%%%%%%
%

Concerning the electrostatic potential outside the dielectric sphere
(region 1) we know that without dielectric discontinuity (at $r=a$) the
potential would simply be given by
$\frac{q}{4\pi \varepsilon _{0}\varepsilon _{1}|\mathbf{r}-\mathbf{b}|}$.
Making use of the following identity

%%%%%%%%%%%%%%%%%%%%%%%%
%FIG 1
%%%%%%%%%%%%%%%%%%%%%%%%

%%%%%%%%%%%%%%%%%%%%%%
\begin{equation}
\label{Eq.1overR}
\frac{1}{|\mathbf{r}-\mathbf{b}|}=
\sum ^{\infty }_{l=0}\frac{r^l_{<}}{r^{l+1}_{>}}P_l(\cos \theta),
\end{equation}
%%%%%%%%%%%%%%%%%%%%%%
%
where $r_{<}$ ($r_{>}$) is the smaller (larger) of $r$
and $b$, the electrostatic potential $\Phi_1(\mathbf{r})$ in region 1 reads
\cite{Note_Eq_1overR}
%

%%%%%%%%%%%%%%%%%%%%%%
\begin{equation}
\label{Eq.FI-out}
\Phi _{1}(r,\theta) =
\sum ^{\infty }_{l=0}\left[ C_{l}\frac{1}{r^{l+1}} +
\frac{q}{4\pi \varepsilon _{0}\varepsilon _{1}}\frac{r^{l}_{<}}{r^{l+1}_{>}}\right]
P_{l}(\cos \theta),
\end{equation}
%%%%%%%%%%%%%%%%%%%%%%
%
recalling that $\Phi_1(\mathbf{r})$ must be finite at $r\rightarrow \infty$
so that $M_{l}=0$  in Eq. (\ref{Eq.FI-general}).

\subsection{Boundary conditions}

The electrostatic potentials given by Eqs. (\ref{Eq.FI-in}) and
(\ref{Eq.FI-out}) will be univocally determined by applying the proper
boundary conditions that will fix  $A_{l}$ and $C_{l}$.
The boundary conditions are derived
from the full set of Maxwell equations.
The results are that the normal components of the displacement $\mathbf{D}$
and the tangential components of $\mathbf{E}$ on either side of the spherical
interface at $r=a$ satisfy

%%%%%%%%%%%%%%%%%%%%%%
\begin{equation}
\label{Eq.boundary-cond}
\left\{
\begin{array}{l}
(\mathbf{D}_{1}-\mathbf{D}_{2})\cdot \mathbf{n}_{12} =  0 \\
(\mathbf{E}_{1}-\mathbf{E}_{2})\times \mathbf{n}_{12} =  0
\end{array}
\right.
\end{equation}
%%%%%%%%%%%%%%%%%%%%%%
%
where $\mathbf{n}_{12}=\mathbf{r}/r$ is a unit normal vector
to the surface directed from region 2 to region 1 (see Fig. \ref{fig.setup-point}).
Within the framework of the linear response theory we have
$\mathbf{D}=\varepsilon \mathbf{E}$.
Combining Eqs. (\ref{Eq.FI-in}) and (\ref{Eq.FI-out}) with Eq.
(\ref{Eq.boundary-cond}) and noting that $\mathbf{E}=-\nabla \Phi$,
it follows that

%%%%%%%%%%%%%%%%%%%%%%
\begin{equation}
\label{Eq.Leg-coeff-EQS}
\left\{
\begin{array}{lll}
\varepsilon_2A_{l}la^{l-1} & = &
\varepsilon_1\left[ -C_{l} \displaystyle\frac{l+1}{a^{l+2}}+
\frac{q}{4\pi \varepsilon_0\varepsilon_1}\frac{la^{l-1}}{b^{l+1}}\right] \\ \\
A_{l}a^{l} & = &
C_{l}\displaystyle\frac{1}{a^{l+1}}+
\frac{q}{4\pi \varepsilon _{0}\varepsilon _{1}}\frac{a^{l}}{b^{l+1}}
\end{array}
\right.
\end{equation}
%%%%%%%%%%%%%%%%%%%%%%
%
This set of two equations {[}Eq. (\ref{Eq.Leg-coeff-EQS}){]} can
be readily solved to yield the Legendre coefficients $A_{l}$
and $C_{l}$:

%%%%%%%%%%%%%%%%%%%%%%
\begin{equation}
\label{Eq.Leg-coeff}
\left\{
\begin{array}{lll}
A_{l} & = & \displaystyle \frac{q}{4\pi \varepsilon_0\varepsilon_1}
\frac{1}{b^{l+1}}
\frac{\varepsilon _{1}(2l+1)}{\varepsilon _{1}(l+1)+\varepsilon_2 l}\\ \\
C_{l} & = & \displaystyle \frac{q}{4\pi \varepsilon_0\varepsilon_1}
\frac{a^{2l+1}}{b^{l+1}}
\frac{(\varepsilon_1-\varepsilon_2)l}{\varepsilon_1(l+1)+\varepsilon_2l}
\end{array}
\right.
\end{equation}
%%%%%%%%%%%%%%%%%%%%%%
%
and hence

%%%%%%%%%%%%%%%%%%%%%%
\begin{eqnarray}
\label{Eq.FI-out-FULL}
%\begin{array}{lll}
\Phi_1(r,\theta) & = & \frac{q}{4\pi \varepsilon_0\varepsilon_1}
\times \left[ \frac{1}{|\mathbf{r}-\mathbf{b}|}   \right. \nonumber \\
&& + \left. \sum ^{\infty }_{l=1}\frac{a^{2l+1}}{b^{l+1}}
\frac{(\varepsilon _{1}-\varepsilon _{2})l}{\varepsilon_1(l+1)+\varepsilon_2 l}
\frac{1}{r^{l+1}}P_l(\cos \theta)
\right].  \nonumber
\\
\end{eqnarray}
%%%%%%%%%%%%%%%%%%%%%%
%

The physical interpretation of Eq. (\ref{Eq.FI-out-FULL}) is straightforward.
The first term represents the usual electrostatic potential (without image forces)
generated by $q$  and the second term can be referred to as the electrostatic
potential due to ``image charges'' stemming from the dielectric discontinuity.
%In terms of many-body physics the second term can be seen as a \textit{non-central}
%field (except for $|\cos \theta |=1$) generated by an \textit{effective}
%particle located at the center of the dielectric sphere.
As expected, the strength of the image force is strongly governed by
the jump $\Delta \varepsilon$ in the dielectric constant defined as

%%%%%%%%%%%%%%%%%%%%%%
\begin{equation}
\label{Eq.D-eps12}
\Delta \varepsilon =\varepsilon_1-\varepsilon_2.
\end{equation}
%%%%%%%%%%%%%%%%%%%%%%
%
In particular, one can anticipate and state that the interaction between the microion $q$
and the dielectric particle (i.e., the \textit{self-image interaction})
is \textit{repulsive} for $\Delta \varepsilon >0$ (i.e., $\varepsilon_1>\varepsilon_2$)
and \textit{attractive} for $\Delta \varepsilon <0$ (i.e., $\varepsilon_1<\varepsilon_2$)
as it is also the case in planar geometry.

One can show that Eq. (\ref{Eq.FI-out-FULL}) can also be written
as follows (see e.g., Ref.\cite{Iversen_98} and references therein)

%%%%%%%%%%%%%%%%%%%%%%
\begin{eqnarray}
\label{Eq.FI-out-IMG}
\Phi_1(r,\theta ) & = & \frac{q}{4\pi
\varepsilon_0\varepsilon_1}
\left\{
\frac{1}{|\mathbf{r}-\mathbf{b}|} +
\frac{\varepsilon_1-\varepsilon_2}
{\varepsilon_1+\varepsilon_2}\frac{1}{a} \right. \nonumber \\
&& \times \left[ \frac{u}{|\mathbf{r}-\mathbf{u}|} \right. \nonumber \\
&& \left. \left. -\frac{\varepsilon_1}{\varepsilon_1+\varepsilon_2}
\int^u_0\frac{(u/x)^{\varepsilon_2/(\varepsilon_1+\varepsilon_2)}}
{|\mathbf{r}-\mathbf{x}|} dx \right]
\right\}, \nonumber \\
\end{eqnarray}
%%%%%%%%%%%%%%%%%%%%%%
%
where $\mathbf{u}=\mathbf{b}a^{2}/b^{2}$ (see Fig. \ref{fig.setup-point}).
\cite{Note_metal}
In this formalism the geometrical structure of the image charges is transparent and
it is specified by the second main term (between brackets) of Eq. (\ref{Eq.FI-out-IMG}).
More precisely, one has to deal with an \textit{infinite manifold}
of image charges distributed along the oriented segment $\mathbf{u}$
that electrically compensates the image point-charge $q_{im}$
located at $\mathbf{u}$ and whose magnitude is given by

%%%%%%%%%%%%%%%%%%%%%%
\begin{equation}
\label{Eq.Qim}
q_{im}=q\frac{\varepsilon_1-\varepsilon_2}{\varepsilon_1+\varepsilon_2}\frac{a}{b}.
\end{equation}
%%%%%%%%%%%%%%%%%%%%%%
%

\subsection{Polarization charge}

It is important to know the surface distribution of the induced charge
on the spherical interface.
In the bulk (i. e., in region 1 or 2) we have a zero volume
density of polarization charge ($\rho _{pol}$) since
$\rho _{pol}=\varepsilon _{0}\nabla \cdot \mathbf{E}=-\nabla \cdot \mathbf{P}=0$
(except at $\mathbf{r}=\mathbf{b}$).
At the interface ($r=a$) the surface density of polarization charge
$\sigma ^{(sph)}_{pol}$ is given by

%%%%%%%%%%%%%%%%%%%%%%
\begin{equation}
\label{Eq.pol-charge}
\sigma ^{(sph)}_{pol}=-(\mathbf{P}_{1}-\mathbf{P}_{2})\cdot \mathbf{n}_{12},
\end{equation}
%%%%%%%%%%%%%%%%%%%%%%
%
where

%%%%%%%%%%%%%%%%%%%%%%
\begin{equation}
\label{Eq.P}
\left\{
\begin{array}{c}
\mathbf{P}_1=
\varepsilon_0(\varepsilon_1-1)\mathbf{E}_1=-\varepsilon_0(\varepsilon_1-1)\nabla\Phi_1\\
\mathbf{P}_{2}=
\varepsilon_0(\varepsilon_2-1)\mathbf{E}_2=-\varepsilon_0(\varepsilon_2-1)\nabla \Phi_2
\end{array}
\right.
\end{equation}
%%%%%%%%%%%%%%%%%%%%%%
%
are the polarizations in region 1 and 2, respectively.
Using Eqs. (\ref{Eq.FI-in}), (\ref{Eq.FI-out}), (\ref{Eq.Leg-coeff}),
(\ref{Eq.pol-charge}) and (\ref{Eq.P}), the final expression of
$\sigma ^{(sph)}_{pol}$  reads

%%%%%%%%%%%%%%%%%%%%%%
\begin{eqnarray}
\label{Eq.pol-charge-FULL}
\sigma ^{(sph)}_{pol}(\cos \theta ) & = &
\frac{q}{4\pi \varepsilon_1b^{2}}
\sum ^{\infty }_{l=1}
\left( \frac{a}{b}\right) ^{l-1}(2l+1)l \nonumber \\
&& \times \frac{\varepsilon_1-\varepsilon_2}
{\varepsilon_1(l+1)+\varepsilon_2l}
P_{l}(\cos \theta ).
\end{eqnarray}
%%%%%%%%%%%%%%%%%%%%%%
%
The net charge of polarization
$Q^{(sph)}_{pol}=\int^{1}_{-1}2\pi a^{2}\sigma ^{(sph)}_{pol}(\cos\theta)d(\cos\theta)$
is zero, \cite{Note_zero_monopole}
meaning that there is \textit{no monopole} contribution as it should be.

The critical angle $\theta^*$ where $\sigma^{(sph)} _{pol}$ changes
sign is given by the geometrical condition

%%%%%%%%%%%%%%%%%%%%%%
\begin{equation}
\label{Eq.E*n}
\left\{ \begin{array}{c}
\mathbf{E}_{1}(r=a,\theta ^{*})\; \bot \; \mathbf{n}_{12}\\
\mathbf{E}_{2}(r=a,\theta ^{*})\; \bot \; \mathbf{n}_{12}
\end{array}\right.
\end{equation}
%%%%%%%%%%%%%%%%%%%%%%
%
which is the orthogonality condition at the interface between the (inner
and outer) electric field and $\mathbf{n}_{12}$. In terms of
Legendre polynomials, Eq. (\ref{Eq.E*n}) can be equivalently
written as

%%%%%%%%%%%%%%%%%%%%%%
\begin{equation}
\label{Eq.teta*}
\sum ^{\infty }_{l=1}
\left( \frac{a}{b}\right) ^{l-1}(2l+1)l
\frac{\varepsilon_1-\varepsilon_2}{\varepsilon_1(l+1)+\varepsilon_2l}
P_{l}(\cos \theta ^{*})=0,
\end{equation}
%%%%%%%%%%%%%%%%%%%%%%
%
where Eq. (\ref{Eq.pol-charge-FULL}) was used.
Two limiting cases can be easily described:
(i) for $b/a\gg 1$ we have $\theta ^{*}\rightarrow \pi /2$
{[}recalling that $P_{1}(\cos \theta )=\cos \theta${]} and
(ii) for $b/a\rightarrow 1$ we have $\theta ^{*}\rightarrow 0$.
In general, $\theta ^{*}$ increases with  $b$ and it is a complicated
function of $b/a$, $\varepsilon_1$ and $\varepsilon_2$.

\begin{figure}[t]
\includegraphics[width = 8.0 cm]{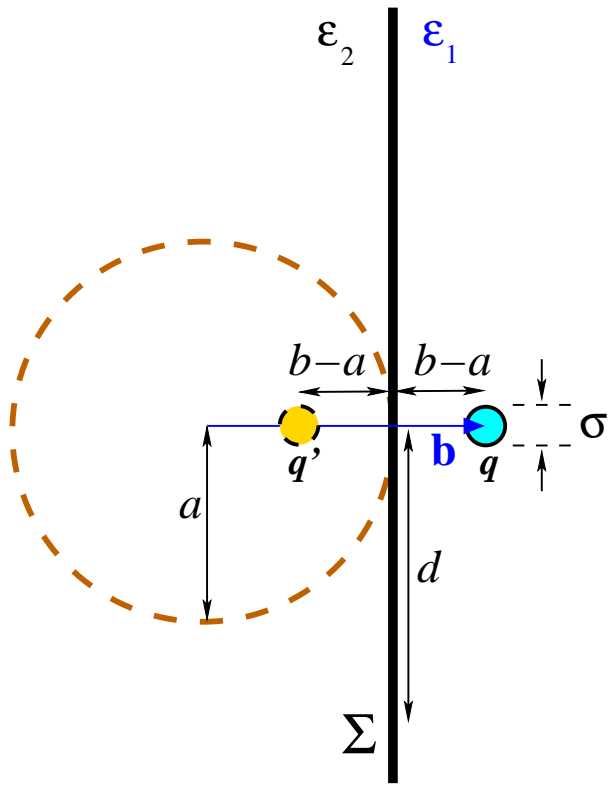}
\caption
{
Model for a microion ($q$) near a planar interface ($\Sigma$)
separating the two infinite media characterized by
$\varepsilon_1$ and $\varepsilon_2$.
The imaginary spherical dielectric of radius $a$ is shown for
 geometrical comparison with the setup of Fig. \ref{fig.setup-point}.
This is a two-dimensional representation of the
three-dimensional system.
}
\label{fig.setup-plan}
\end{figure}

For a \textit{planar} interface, the surface density of polarization
charge $\sigma ^{(plan)}_{pol}(d)$ is given by \cite{Jackson_book_1975}

%%%%%%%%%%%%%%%%%%%%%%
\begin{equation}
\label{Eq.pol-charge-PLAN}
\sigma ^{(plan)}_{pol}(d)=
\frac{q}{2\pi \varepsilon_1}
\frac{\varepsilon_1-\varepsilon_2}{\varepsilon_1+\varepsilon_2}
\frac{b-a}{\left[ (b-a)^{2}+d^{2})\right] ^{3/2}},
\end{equation}
%%%%%%%%%%%%%%%%%%%%%%
%
where $d=\sqrt{x^{2}+y^{2}}$ is the radial distance (in cylindrical
coordinates system) belonging to the planar interface (see Fig. \ref{fig.setup-plan}).
Equation (\ref{Eq.pol-charge-PLAN}) demonstrates that $\sigma ^{(plan)}_{pol}(d)$
\textit{never} changes sign {[}as can also be deduced from simple
geometrical considerations - Eq. (\ref{Eq.E*n}){]} in contrast with
the spherical interface. The total charge of polarization $Q^{(plan)}_{pol}$
is obtained by direct integration of $\sigma ^{(plan)}_{pol}(d)$
and its expression is given by

%%%%%%%%%%%%%%%%%%%%%%
\begin{equation}
\label{Eq.Qpol-PLAN}
Q^{(plan)}_{pol}=\frac{q'}{\varepsilon_1},
\end{equation}
%%%%%%%%%%%%%%%%%%%%%%
%
where

%%%%%%%%%%%%%%%%%%%%%%
\begin{equation}
\label{Eq.q'}
q'=q\frac{\varepsilon_1-\varepsilon_2}{\varepsilon_1+\varepsilon_2}
\end{equation}
%%%%%%%%%%%%%%%%%%%%%%
%
is the \textit{unique} image charge located at the mirror position of $q$
(see Fig. \ref{fig.setup-plan}).
This \textit{non-zero} monopolar contribution for the planar interface
involves a \textit{stronger} and \textit{longer ranged} self-image
interaction.

%%%%%%%%%%%%%%%%%%%%%%%%
%FIG 2
%%%%%%%%%%%%%%%%%%%%%%%%

\subsection{Application to colloidal systems\label{Sec.Application-colloid}}

So far we treat in a rather general manner the physics of a point
charge near a spherical dielectric interface. We now
would like to apply the above theory to colloidal systems. In the
remaining of this paper we suppose that region 1 corresponds to water,
so that we take $\varepsilon_1=80$ corresponding to the water
dielectric constant at room temperature. To characterize the low permittivity
of the colloid we consider here $\varepsilon_2=2$ so that
$\Delta \varepsilon =78$. The little ion carries a charge $q=Ze$
where $e$ stands for the elementary charge and $Z$ for its
valence, and has a diameter $\sigma $.
An important quantity is

%%%%%%%%%%%%%%%%%%%%%%
\begin{equation}
\label{Eq.r0}
r_{0}=a+\frac{\sigma }{2}
\end{equation}
%%%%%%%%%%%%%%%%%%%%%%
%
being the center-center distance of closest approach between
the colloid and the microion $q$.

\subsubsection{Induced surface charge\label{Sec.induced-charge}}

It is helpful to have a precise representation of the polar profile
of $\sigma ^{(sph)}_{pol}(\theta)$ in order to get a clear understanding
of the source of the image forces.
Although at first glance such a study should belong to standard
electrostatics we are not aware of any data in the literature
that treats this crucial aspect.

%%%%%%%%%%%%%%%%%%%%%%%%%
%FIG 3
\begin{figure}
\includegraphics[width = 8.0 cm]{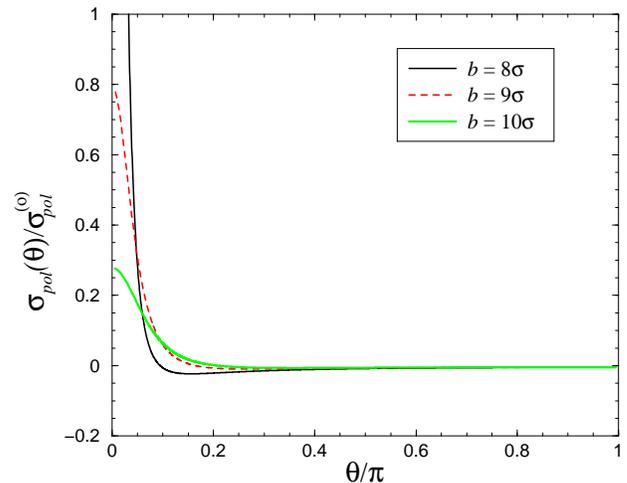}
\caption
{
Polar profile of the surface density of polarization charge
$\sigma^{(sph)} _{pol}(\theta)$ in units of
$\sigma ^{(0)}_{pol}=\frac{q}{4\pi \varepsilon_1\sigma ^{2}}$
for different radial distances $b $ of the excess
charge $q $ with $\varepsilon_1=80$,
$\varepsilon_2=2 $ and $a=7.5\sigma$.
}
\label{fig.Qpol}
\end{figure}
%%%%%%%%%%%%%%%%%%%%%%%%%

The numerical computation of Eq. (\ref{Eq.pol-charge-FULL}) was performed
using a cutoff $l_{max}=300$ in the Legendre space leading to high accuracy.
\cite{Note_convergence_pol}
The plot of $\sigma ^{(sph)}_{pol}(\theta )$ for $a=7.5\sigma$ and
$b/\sigma =8$, 9 and 10 can be found in Fig. \ref{fig.Qpol}.
One can clearly observe that $\sigma ^{(sph)}_{pol}(\theta)$
is strongly \textit{inhomogeneous}.
For small $\theta$, $\sigma ^{(sph)}_{pol}(\theta )$
is highly positive (i.e., it carries the same charge sign as $q$)
and decreases abruptly.
The angle $\theta ^{*}$ {[}given by Eq. (\ref{Eq.teta*}){]} where
$\sigma ^{(sph)}_{pol}(\theta )$ changes sign is
$16.9^{\circ}$, $29.5^{\circ }$ and $37.4^{\circ }$
for $b/\sigma =8$, 9 and 10, respectively.
In parallel, by increasing $b$ one drastically decreases
the magnitude  as well as the inhomogeneity of $\sigma ^{(sph)}_{pol}(\theta)$.
Recall that for $b/a \gg 1$ we have
$\sigma ^{(sph)}_{pol}(\theta)\sim\cos\theta$.

%%%%%%%%%%%%%%%%%%%%%%%%%%%%%%%%%%
% TABLE 1
\begin{table}[b]
\caption
{
Numerical values of $\sigma ^{(sph)}_{pol}(\theta =0)$ and
$\sigma ^{(plan)}_{pol}(d=0)$ in units of
$\frac{q}{4\pi \varepsilon_1\sigma ^{2}}$
as a function of $b$.
The corresponding profiles of  $\sigma ^{(sph)}_{pol}(\theta)$
can be found in Fig.~\ref{fig.Qpol}.
}
\begin{ruledtabular}
\begin{tabular}{lcc}
$b/\sigma$ &
$\sigma ^{(sph)}_{pol}(\theta =0)$ &
$\sigma ^{(plan)}_{pol}(d=0)$  \\
\hline
8&
7.41&
7.61\\
9&
0.794&
0.846\\
10&
0.278&
0.304\\
\end{tabular}
\label{tab.pol}
\end{ruledtabular}
\end{table}
%%%%%%%%%%%%%%%%%%%%%%%%%%%%%%%%%%

It is insightful to compare $\sigma ^{(sph)}_{pol}(\theta=0)$
with  $\sigma ^{(plan)}_{pol}(d=0)$ [computed from Eq.~\eqref{Eq.pol-charge-PLAN}]
since both quantities give the maximum of $\sigma ^{(sph)}_{pol}(\theta)$ and
$\sigma ^{(plan)}_{pol}(d)$, respectively.
The corresponding numerical values are gathered in Table~\ref{tab.pol}.
The  values found at finite curvature are very similar
to those of zero one.
The fact that $\sigma ^{(sph)}_{pol}(\theta=0)$ is systematically
smaller than $\sigma ^{(plan)}_{pol}(d=0)$ is consistent with
the idea that in spherical geometry we have the presence of opposite
image charges.
Nevertheless, for sufficiently large $a$ one should recover the planar case.
%All these features will be important to understand (i) the underlying
%physics of our simulation results involving \textit{many} microions
%and (ii) more generally the effects of image charges in spherical
%geometry.

\subsubsection{Self-image interaction\label{Sec.Vself}}

We now compute the potential of interaction between the microion $q$
and the dielectric particle or, in terms of image forces,
the potential of self-image interaction.
This is the work done in bringing the microion from infinity to
its position $\mathbf{b}$, and it is equal to the \textit{half}-product
of $q$ and the second term of $\Phi_1(r=b)$ given by Eq. \eqref{Eq.FI-out-FULL}.
In that case we have $\mathbf{r}=\mathbf{b}$ (see Fig. \ref{fig.setup-point}),
so that $\theta =0$ and therefore $P_{l}[\cos (\theta =0)]=1$.
In order to normalize the energy with $k_{B}T$ we introduce the Bjerrum
length $l_{B}=e^{2}/(4\pi \varepsilon _{0}\varepsilon_1k_{B}T)$
which is $7.14$ \AA\ for water at $T=298$ K.
By choosing $\sigma =3.57$ \AA\ we have $l_{B}=2\sigma$.
The potential of self-image interaction $V^{(sph)}_{self}(b)$
is then given by

%%%%%%%%%%%%%%%%%%%%%%
\begin{equation}
\label{Eq.Vr-sphere}
V^{(sph)}_{self}(b) =
\frac{1}{2}k_B T l_B \frac{Z^2}{b}
\sum ^{\infty }_{l=1}
\left( \frac{a}{b}\right) ^{2l+1}
\frac{(\varepsilon_1-\varepsilon_2)l}{\varepsilon_1(l+1)+\varepsilon_2l}.
\end{equation}
%%%%%%%%%%%%%%%%%%%%%%
%
Equation (\ref{Eq.Vr-sphere}) shows that the typical interaction
range scales like $1/b^{4}$ and therefore it is
\textit{short-ranged}. \cite{Note_short_range} Note that it is
fully equivalent to compute $V^{(sph)}_{self}(b)$ from the surface
polarization charges as follows

%%%%%%%%%%%%%%%%%%%%%%
\begin{equation}
\label{Eq.Vself-IMG}
V^{(sph)}_{self}(b)=
\frac{1}{2}\frac{1}{4\pi \varepsilon _{0}}
\int ^{1}_{-1}2\pi a^{2}q
\frac{\sigma ^{(sph)}_{pol}(\cos \theta )}{|\mathbf{r}_{a}-\mathbf{b}|}
d(\cos \theta ),
\end{equation}
%%%%%%%%%%%%%%%%%%%%%%
%
where $\mathbf{r}_{a}$ is the radial vector of magnitude $|\mathbf{r}_{a}|=a$
and $\sigma ^{(sph)}_{pol}(\cos \theta )$ is given by Eq. (\ref{Eq.pol-charge-FULL}).

It is insightful to compare the  potential of self-image interaction obtained
in spherical geometry with that, $V^{(plan)}_{self}(b-a)$, obtained in planar geometry.
The setup for a planar interface is sketched in Fig. \ref{fig.setup-plan}.
In this situation the analytical expression of $V^{(plan)}_{self}(b-a)$
is simply given by

%%%%%%%%%%%%%%%%%%%%%%
\begin{equation}
\label{Eq.Vr-plan}
V^{(plan)}_{self}(b-a)=
\frac{1}{2}k_{B}Tl_{B}Z^{2}
\frac{\varepsilon_1-\varepsilon_2}{\varepsilon_1+\varepsilon_2}
\frac{1}{2(b-a)}.
\end{equation}
%%%%%%%%%%%%%%%%%%%%%%

\begin{figure}[t]
\includegraphics[width = 8.0 cm]{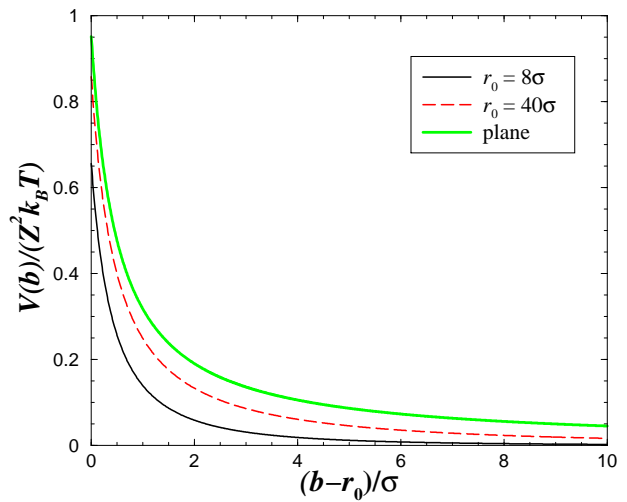}
\caption
{
Potential of self-image interaction for a microion ($q=Ze$)
in spherical and planar geometries with $\varepsilon_1=80 $
and $\varepsilon_2=2$.
}
\label{fig.Vr}
\end{figure}

Profiles of $V^{(sph)}_{self}(b)$ (for two colloidal radii) and
$V^{(plan)}_{self}(r)$ are reported in Fig. \ref{fig.Vr}. Since
in both (planar and spherical) cases the potential of interaction
diverges at the interface, we only show results from $r>r_{0}$ as
it is the case in experimental systems.
The numerical computation of Eq. (\ref{Eq.Vr-sphere})
was performed using the formalism of Eq.~\eqref{Eq.FI-out-IMG}
allowing an arbitrary precision. 
\cite{Note_comparison_2methods}

%%%%%%%%%%%%%%%%%%%%%%%%
%FIG 4
%%%%%%%%%%%%%%%%%%%%%%%%

Figure \ref{fig.Vr} clearly shows that the self-image interaction
is weaker (the higher the curvature) with a spherical interface than
with a planar one. In particular, at contact we have
$V^{(sph)}_{self}(r_{0}=8\sigma )=0.66Z^{2}k_{B}T$
and $V^{(sph)}_{self}(r_{0}=40\sigma )=0.86Z^{2}k_{B}T$ for the
spherical interface and $V^{(plan)}_{self}(\sigma/2)=0.95Z^{2}k_{B}T$
for the planar one.
These features can be physically explained in terms of polarization charges.
In the contact region (i. e., for small $\theta$ - see Fig. \ref{fig.setup-plan})
we know that the surface polarization charge is quasi-identical on
both spherical and planar interfaces. However, for \textit{finite}
curvature we also know that $\sigma^{(sph)}_{pol}$ changes sign above
$\theta ^{*}$ and in the present case $\sigma^{(sph)}_{pol}$ gets
\textit{oppositely} charged to $q$. This latter effect is the
main cause that leads to a weaker self-image interaction for spherical
interfaces.
Nevertheless, by increasing $a$ (i.e., reducing the curvature)
one approaches the planar case as expected (see Fig. \ref{fig.Vr}).
Physically, this means that the contribution of the negative
polarization charges (lying at $\theta >\theta ^{*}$) to the self-image
interaction [Eq. (\ref{Eq.Vself-IMG})] becomes negligible
for sufficiently large colloidal radius.

\subsubsection{Effect of curvature on the contact potential\label{Sec.Curvature}}

It is clear that for sufficiently low curvature one should recover
the planar case as far as the self-image interaction is concerned.
Thus, a natural question that arises is: for which typical colloidal
size are curvature effects relevant?

A suitable observable for this problem is provided by the contact
potential $V^{(sph)}_{self}(b=a+\sigma/2)$.
This quantity is of special interest since it will correspond to the
highest repulsive part of the global interaction between a macroion
(i.e., \textit{charged} macro-particle) and an oppositely charged counterion.
In order to investigate the effect of finite curvature we are going
to compare $V^{(sph)}_{self}(a+\sigma/2)$ to the contact potential
$V^{(plan)}_{self}(b-a=\sigma/2)$ obtained with a planar interface.

\begin{figure}[b]
\includegraphics[width = 8.0 cm]{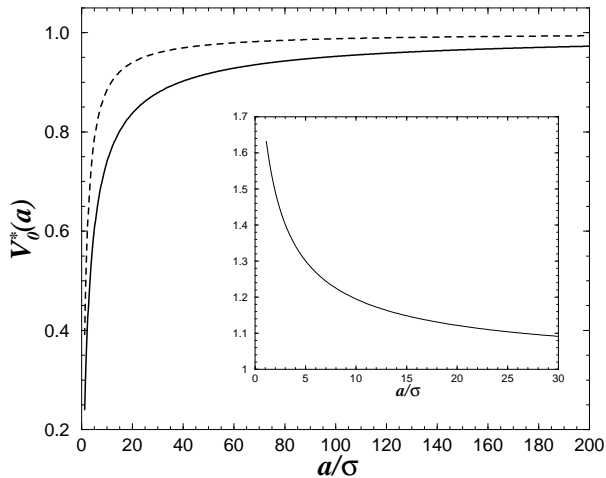}
\caption
{
Reduced contact potential $V^{*}_{0}(a) $ as
a function of the colloidal radius $a $ with $\varepsilon_1=80 $,
$\varepsilon_2=2$. The limit value of unity
corresponds to the planar interface. The solid line is the exact contact
potential $V^{*}_{0}(a) $ and the dashed one is
the contact potential $\widetilde{V}^{*}_{0}(a)$
obtained with the two-image charge approximation used by Linse.
\cite{Linse_JPC_1986}
The insert shows the ratio $\widetilde{V}^{*}_{0}(a)/V^{*}_{0}(a)$.
}
\label{fig.V0}
\end{figure}

The plot of the normalized contact potential $V^{*}_{0}(a)$ defined
as

%%%%%%%%%%%%%%%%%%%%%%
\begin{equation}
\label{Eq.V0*}
V^{*}_{0}(a)=\frac{V^{(sph)}_{self}(a+\frac{\sigma}{2})}
{V^{(plan)}_{self}(\frac{\sigma}{2})}
\end{equation}
%%%%%%%%%%%%%%%%%%%%%%
%
can be found in Fig. \ref{fig.V0}. For the sake of numerical
stability we used the formalism of Eq. (\ref{Eq.FI-out-IMG})
allowing an arbitrary precision. \cite{Note_comparison_2methods}
Figure \ref{fig.V0} shows that for $a/\sigma$ larger than about
$100$ the contact potential is close to that of the planar
interface (less than $5\%$ difference). This length scale
typically corresponds to {}``true'' colloidal systems ($\sim$ 100
nm). Therefore, in the dilute regime where the self-image
interaction is dominant (i.e. lateral microion-microion
correlations are negligible), large-sized colloidal particles can
be reasonably approximated by planar interfaces as far as the
modeling of the self-image interaction is concerned. On the other
hand, for $a/\sigma$ smaller than about $20$ the contact potential
varies rapidly and therefore it is strongly dependent on the
curvature. This length scale typically corresponds to micellar
systems ($\sim$ 10 nm).
%In this latter case, it is clear that a detailed specific study is
%required. This is the purpose of our forthcoming simulations which
%will treat \textit{explicitly} the difficult many-body problem taking
%place in charged micellar systems.

%%%%%%%%%%%%%%%%%%%%%%%%%%%%%%%%%%%%%%%%
%FIG 5
%%%%%%%%%%%%%%%%%%%%%%%%%%%%%%%%%%%%%%%%

In this respect, Linse \cite{Linse_JPC_1986} used  an uncontrolled
approximation where he replaced the (exact) infinite manifold of image
 charges {[}entering Eq. (\ref{Eq.FI-out-IMG}){]} of total charge $-q_{im}$
by a single image point-charge $-q_{im}$ {[}given by Eq. (\ref{Eq.Qim}){]}
located at the center of the sphere.\cite{Note_Linse_a}
Doing so the setup of image charges consists of a (two point-charge) dipole
$\mathbf{p}_{im}=q_{im}\mathbf{u}$, and the corresponding contact
potential $\widetilde{V}^{(sph)}_{self}(r_0)$
reads

%%%%%%%%%%%%%%%%%%%%%%
\begin{equation}
\label{Eq.V0-sphere-LINSE}
\widetilde{V}^{(sph)}_{self}(b=r_0) =
k_{B}Tl_{B}\frac{Z^2}{2}
\frac{\varepsilon_1-\varepsilon_2}{\varepsilon_1+\varepsilon_2}
\frac{a}{r_{0}}\left[ \frac{1}{r_{0}-u}-\frac{1}{r_{0}}\right].
\end{equation}
%%%%%%%%%%%%%%%%%%%%%%
%
The plot of

\begin{equation}
\label{Eq.V0*-Linse}
\widetilde{V}^*_0(a) =
\frac{\widetilde{V}^{(sph)}_{0}(a+\frac{\sigma}{2})}
{V^{(plan)}_{self}(\frac{\sigma}{2})}
\end{equation}
can also be found in Fig. \ref{fig.V0}. It shows that the two-image charge
approximation used by Linse is only valid for very low curvature (i.e.,
close to the planar case) and may strongly overestimate the self-image
repulsion as expected by its inherent construction.
\cite{Note_Linse_b}
In his MC simulations, Linse \cite{Linse_JPC_1986} investigated
micelles of radius $12-18$ \AA\ (i.e, $a/\sigma \sim 3.5-5$)
leading to errors as large as $40\%$ (see insert of Fig. \ref{fig.V0}).
This proves that this ansatz is
unsuitable to determine the  self-image interaction in this regime, which is the
source of the image forces.
%Our simulations will show that, especially for micellar systems, the image
%forces in many-counterion systems become only relevant when the self-image
%interaction is dominant.

\subsubsection{Charged colloid\label{Sec.Vm}}

As a last theoretical result, we consider the interaction between
(a single counterion) $q$  and a negatively \textit{charged} dielectric sphere.
The procedure is completely similar to the neutral colloid case, and
we now apply the principle of superposition to take into account the
additional potential due to a central charge $Q_m=-Z_me$.
The (global) macroion-counterion potential of interaction $V_{m}^{}(b)$ reads

%%%%%%%%%%%%%%%%%%%%%%
\begin{equation}
\label{Eq.Vm}
V_{m}(b)=-k_{B}Tl_{B}\frac{Z_mZ}{b}+V^{(sph)}_{self}(b)
\end{equation}
%%%%%%%%%%%%%%%%%%%%%%
%
where $V^{(sph)}_{self}(b)$ is given by Eq. (\ref{Eq.Vr-sphere}),
and hence

%%%%%%%%%%%%%%%%%%%%%%
\begin{eqnarray}
\label{Eq.Vm-FULL}
V_{m}^{}(b) & = & k_{B}T\frac{l_{B}}{b}Z^{2}
\left[ -\frac{Z_m}{Z} \right. \nonumber \\
&& +
\left. \frac{1}{2}\sum ^{\infty }_{l=1}\left( \frac{a}{b}\right) ^{2l+1}
\frac{(\varepsilon_1-\varepsilon_2)l}{\varepsilon_1(l+1)+\varepsilon_2l}\right].
\end{eqnarray}
%%%%%%%%%%%%%%%%%%%%%%
%
%%%%%%%%%%%%%%%%%%%%%%%%%%%%%%%%%%%%%%%%%
%FIG 6
\begin{figure}
\includegraphics[width = 8.0 cm]{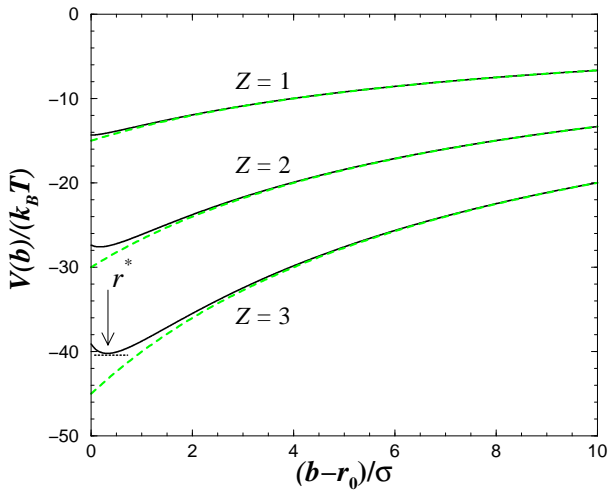}
\caption
{
Global macroion-counterion potential of interaction (solid lines) with $Z_m=60$,
$r_0=8\sigma$, $\varepsilon_1=80$ and $\varepsilon_2=2$.
The values of the corresponding minima $r^{*} $ can be found in
Table \ref{tab.r*}.
The dashed lines correspond to the usual electrostatic potential of
interaction without image forces (i.e., $\Delta\varepsilon=0$).
}
\label{fig.Vm}
\end{figure}
%%%%%%%%%%%%%%%%%%%%%%%%%%%%%%%%%%%%%%%%%

Profiles of $V_m(b)$ for $Z_m=60$, $r_0=8\sigma$, $\varepsilon_2=2$
and $Z=1$, 2 and 3 are reported in Fig. \ref{fig.Vm}.
An important result is the occurrence of a \textit{minimum}
in $V_m(b)$ whose depth and position $r^*$ increase with increasing $Z$.
This is due to the purely \textit{repulsive} self-image interaction
which scales like $Z^{2}$, whereas the direct \textit{attractive}
Coulomb macroion-microion interaction scales like $Z$ (at fixed $Z_m $).
Nevertheless the occurrence of a minimum is strongly dictated by the
ratio $Z_m/Z$ [see Eq. (\ref{Eq.Vm-FULL})]. For high value of $Z_m/Z$,
$|V_{m}^{}(b)|$ is maximal for $b=r_0 $ (only \textit{attraction}
occurs) and for small $Z_m/Z$ one recovers the neutral colloid
case where only \textit{repulsion} occurs.
Of course the same qualitatively happens for charged plates.
\cite{Note_plate}
The values of $r^{*} $ minimizing
$V_{m}^{}(b) $ (with $b>r_0 $) are given in Table \ref{tab.r*}.
The quantity $r^{*} $ will be useful to discuss our simulation
results that concern \textit{many} counterions and where we also have
the same macroion bare charge ($Z_m=60$).

Keep in mind that all our results above concern a single microion.
When \textit{many} counterions come into play, other important
effects might appear in principle.
In particular, when the number of  counterions near the
macroion surface is very large the image forces are practically
canceled by symmetry reason.
\cite{Holm_private}
Clearly, by approaching the (perfect) spherical symmetry one asymptotically
cancels the polarization charges everywhere on the macroion surface.
%because all positive polarization charges get (asymptotically)
%exactly compensated by the negative ones.
This point shows that the discrete nature of the counterions is crucial
for the existence of image charges in spherical geometry.
\cite{Note_discretization}
In planar geometry the situation is radically different, where one gets
an amplified image force upon increasing the number of ``surface'' counterions.

%%%%%%%%%%%%%%%%%%%%%%%%%%%%%%%%%%
% TABLE 2
\begin{table}[b]
\caption
{
Theoretical values of $r^*$ minimizing the macroion-counterion potential
of interaction (with $Z_m=60$, $\varepsilon_1=80$, $\varepsilon_2=2$
 and $r_0=8\sigma$).
The corresponding profiles can be found in Fig. \ref{fig.Vm}.
}
\begin{ruledtabular}
\begin{tabular}{lccc}
$\varepsilon_2$ &
$\Delta \varepsilon$ &
$ Z$ &
$(r^*-r_0)/\sigma$\\
\hline
2&
78&
1&
0\\
2&
78&
2&
0.17\\
2&
78&
3&
0.32\\
\end{tabular}
\label{tab.r*}
\end{ruledtabular}
\end{table}
%%%%%%%%%%%%%%%%%%%%%%%%%%%%%%%%%%

%%%%%%%%%%%%%%%%%%%%%%%%%%%%%%%%%%%%%%%%%%%%%%%%%%%%%
\section{Monte Carlo simulation \label{Sec.MC}}
%%%%%%%%%%%%%%%%%%%%%%%%%%%%%%%%%%%%%%%%%%%%%%%%%%%%%

Standard canonical MC simulations following the Metropolis scheme
were used. \cite{Metropolis_JCP_1953,Allen_book_1987} The system
we consider is similar to those studied in previous works.
\cite{Messina_PRL_2000,Messina_EPL_2000,Messina_PRE_2001}
%and its setup is sketched in Fig. \ref{fig.setup-MC}.
It is made up of two types of charged hard spheres: (i) a macroion
of radius $a$ with a bare charge $Q_{m}=-Z_me$ (with $Z_m>0$) and
(ii) small microions (counterions and coions) of diameter $\sigma$
with charge $q=\pm Ze$ to ensure the electroneutrality of the
system. All these ions are confined in an impermeable cell of
radius $R$ and the macroion is held fixed at the center of the
cell.

The dielectric media are modeled as in Sec \ref{Sec.Theory}. The
outer region of the simulation cell is assumed to have the same dielectric
constant $\varepsilon_1$ as the solvent in order to avoid
the appearance of artificial image forces.

The work done in bringing the (real) ions together from infinite
separation gives the interaction energy of the system. The
corresponding Hamiltonian, $U_{tot}$, can  be expressed as

%%%%%%%%%%%%%%%%%%%%%%%%%%%%%%%
\begin{equation}
\label{Eq.MC-U-tot} U_{tot}  =
\sum_i \left[U^{(m)}_i+ \sum_{j>i} U^{(bare)}_{ij} \right]
+ \sum_i\left[U^{(self)}_i + \sum _{j>i}U^{(im)}_{ij}\right].
\end{equation}
%%%%%%%%%%%%%%%%%%%%%%%%%%%%%%%
%
%All these terms entering Eq. (\ref{Eq.MC-U-tot}) are illustrated
%in Fig. \ref{fig.setup-MC}.
%The factor $\frac{1}{2}$ that appears
%(as in planar geometry
%\cite{Gruenberg_JPhysCondM_2000,Torrie_JCP_1982})
%explicitly in $U^{(self)}_{i}$ and implicitly in $U^{(im)}_{ij}$
%avoids double counting

The first two terms in Eq. (\ref{Eq.MC-U-tot}) correspond to the
traditional electrostatic interactions between real charges.
More explicitly,

%%%%%%%%%%%%%%%%%%%%%%%%%%%%%%%
\begin{equation}
\label{Eq.MC-Um}
U^{(m)}_{i}(r_{i})=
\left\{ \begin{array}{l}
\displaystyle \pm l_{B}k_{B}T\frac{Z_mZ}{r_{i}},
\hspace{0.5cm} \textrm{for}~ r_{i}\geq a+\frac{\sigma }{2}, \\ \\
\displaystyle \infty ,
\hspace{2.3cm} \textrm{for}~ r_{i}<a+\frac{\sigma }{2},
\end{array}
\right.
%\qquad \right.
%\begin{array}{l}
%\displaystyle\textrm{for}\, \, r_{i}\geq a+\frac{\sigma }{2},\\
%\displaystyle\textrm{for}\, \, r_{i}<a+\frac{\sigma }{2}.
%\end{array}
\end{equation}
%%%%%%%%%%%%%%%%%%%%%%%%%%%%%%%
%
represents the macroion-microion interaction,
where (+) applies to coions and (-) to counterions,
and

%%%%%%%%%%%%%%%%%%%%%%%%%%%%%%%
\begin{equation}
\label{Eq.MC-Ubare}
U^{(bare)}_{ij}(r_{ij})=\left\{
\begin{array}{l}
\displaystyle \pm l_{B}k_{B}T\frac{Z^{2}}{r_{ij}},
\hspace{0.5cm} \textrm{for}\, \, r_{ij}\geq \sigma,\\
\displaystyle \infty ,
\hspace{2.0cm} \textrm{for}\, \, r_{ij} < \sigma,\\
\end{array}
\right.
%\qquad \right. \begin{array}{l}
%\textrm{for}\, \, r_{ij}\geq \sigma ,\\
%\textrm{for}\, \, r_{ij}<\sigma ,
%\end{array}
\end{equation}
%%%%%%%%%%%%%%%%%%%%%%%%%%%%%%%
%
the pair interaction between  microions
$j$ and $i$ where (+) applies to microions of
the same type and (-) otherwise.

The two last terms in Eq. (\ref{Eq.MC-U-tot})
account for the interaction between images and microions.
The \textit{repulsive} self-image interaction is given by

%%%%%%%%%%%%%%%%%%%%%%%%%%%%%%%
\begin{equation}
\label{Eq.MC-Uself}
U^{(self)}_{i}(r_{i}) =
\left\{ \begin{array}{l}
\displaystyle \frac{1}{2}k_{B}Tl_{B}\frac{Z^{2}}{r_{i}}
\displaystyle\sum ^{l_{max}}_{l=1}
\left(\frac{a}{r_{i}}\right) ^{2l+1} \\ \\
\times
\displaystyle\frac{(\varepsilon_1-\varepsilon_2)l}
{\varepsilon_1(l+1)+\varepsilon_2l},
\hspace{0.5cm} \textrm{for}~ r_{i}\geq a+\frac{\sigma }{2}, \\ \\
\displaystyle \infty,
\hspace{2.7cm} \textrm{for}~ r_{i}<a+\frac{\sigma }{2},
\end{array} \right.
\end{equation}
%%%%%%%%%%%%%%%%%%%%%%%%%%%%%%%
%
where $l_{max}$ is the cutoff in the Legendre space,
and

%%%%%%%%%%%%%%%%%%%%%%%%%%%%%%%
\begin{equation}
\label{Eq.MC-Uim}
U^{(im)}_{ij}(\mathbf{r}_{i},\mathbf{r}_{j})=
\left\{
\begin{array}{l}
\displaystyle \pm l_{B}k_{B}TZ^{2}
\sum ^{l_{max}}_{l=1}\frac{a^{2l+1}}{r_{j}^{l+1}}
\frac{(\varepsilon_1-\varepsilon_2)l}
{\varepsilon_1(l+1)+\varepsilon_2l} \\ \\
\displaystyle \times \frac{1}{r_{i}^{l+1}}P_{l}(\cos \theta ),
\hspace{0.5cm} \textrm{for}~ r_{i}\geq a+\frac{\sigma }{2},\\ \\
\displaystyle \infty ,
\hspace{2.5cm} \textrm{for}~ r_{i}<a+\frac{\sigma }{2},
\end{array}  \right.
\end{equation}
%%%%%%%%%%%%%%%%%%%%%%%%%%%%%%%
%
represents the interaction between microion $i$ and the image
(surface charge induced by)  of microion
$j$, where (+) applies to charges of the same sign {[}and (-) otherwise{]}
and $\theta$ is the angle between $\mathbf{r}_{i}$ and
$\mathbf{r}_{j}$.
It is this term that generates \textit{lateral}
image-counterion correlations.
Due to the symmetry of $U^{(im)}_{ij}$ upon exchanging $ij$ with $ji$ there is
an implicit factor $1/2$ in Eq.~\eqref{Eq.MC-Uim}.

%%%%%%%%%%%%%%%%%%%%%%%%%%%%%%%%%%%%%
%FIG 7
%\begin{figure}[t]
%\includegraphics[width = 8.0 cm]{fig7-JCP.eps}
%\caption
%{
%Simulation model system for a spherical macroion surrounded by small
%salt-microions. The dashed lines indicate typical particles interacting
%with microion $q_{i}$.
%The interaction between $Q$ and $q_{i}$ stands for
%the direct Coulomb macroion-microion interaction.
%The interaction between $q_{j}$ and $q_{i}$
%is the direct Coulomb microion-microion interaction. The fictive {}``image
%charges'' $q'$ are only shown to illustrate
%the presence of interactions due to the dielectric discontinuity
%$\Delta \varepsilon =\varepsilon_1-\varepsilon_2$.
%Within this picture, the interaction between $q_{i}'$
%and $q_{i}$ stands for the self-image interaction,
%and the interaction between $ q_{j}'$ and $ q_{i}$
%stands for the lateral image-microion correlation.
%}
%\label{fig.setup-MC}
%\end{figure}
%%%%%%%%%%%%%%%%%%%%%%%%%%%%%%%%%%%%%

Convergence of the Legendre sums with a relative error of $10^{-6}$
is obtained with the employed value of $l_{max}=100$. 
%as already mentioned in Sec. \ref{Sec.Vself}.
\cite{Note_convergence_Vself}
For the sake of computational efficiency and without loss of accuracy,
we computed  the image-ion interactions on a (very) fine
$(r,\cos \theta)$ grid where the coordinates of the microions were
extrapolated.
The radial distances $r_i$ are discretized over logarithmically equidistant nodes
so that close to the macroion surface the radial resolution is
$0.01\sigma$ and near the simulation wall $0.1\sigma$.
The polar discretization consists of $2000$ equidistant $\cos \theta$-nodes
leading to even smaller lateral resolutions.
The corresponding values of  $U^{(self)}_i(r_i)$ and
$U^{(im)}_{ij}(r_i,r_j,\cos\theta)$ were then  initially stored into tables.
Note that it in principle one could also have used the formalism of
Eq.~\eqref{Eq.FI-out-IMG} to compute the image-ion interactions. However,
at identical numerical accuracy, this method involving a numerical integration
is too time and resource consuming.
%%%%%%%%%%%%%%%%%%%%%%%%%%%%%%%%%%%%%%%%%%%%%%%%%%%%%%%%
% TABLE 3
\begin{table}[b]
\caption{
Model simulation parameters with some fixed values. Apart
from the charge sign, counterions and coions have the same parameters.
}
\label{tab.simu-param}
\begin{ruledtabular}
\begin{tabular}{lc}
 Parameters&
\\
\hline
 $T=298K$&
 room temperature\\
$\varepsilon_1=80$ &
water solvent dielectric constant\\
$\varepsilon_2=2 $&
colloidal dielectric constant\\
 $\Delta \varepsilon =\varepsilon_1-\varepsilon_2 =78$ &
strength of dielectric discontinuity\\
 $Z_m$ &
 macroion valence\\
 $Z $&
 counterion valence\\
$\sigma =3.57$ \AA\ &
 counterion diameter\\
$l_{B}=2\sigma =7.14$ \AA\ &
 Bjerrum length\\
 $a=7.5\sigma$ &
 macroion radius \\
 $r_0=a+\frac{\sigma }{2}=8\sigma$ &
 macroion-counterion distance\\ & of closest approach \\
$R $&
radius of the outer simulation cell\\
\end{tabular}
\end{ruledtabular}
\end{table}
%%%%%%%%%%%%%%%%%%%%%%%%%%%%%%%%%%%%%%%%%%%%%%%%%%%%%%%%

Typical simulation parameters are gathered in Table \ref{tab.simu-param}.
The case $\varepsilon_2=80$ corresponds to the situation where
there is \textit{no} dielectric discontinuity ($\Delta \varepsilon =0$).
%In practice,
%an equilibration of $3\times10^3$ to $2\times10^4$ MC steps per
%particle was necessary for convergence.
Measurements were performed over $10^6$ MC steps per particle.

%%%%%%%%%%%%%%%%%%%%%%%%%%%%%%%%%%%%%%%%%%%%%%%%%%%%%
\section{Simulation results\label{Sec.SIMU-results}}
%%%%%%%%%%%%%%%%%%%%%%%%%%%%%%%%%%%%%%%%%%%%%%%%%%%%%

Here we present our MC simulation results in salt-free environment
as well as in the presence of multivalent salt-ions.
We  essentially study in detail the radial microion distributions
$n_{i}(r)$ around the macroion, which are normalized as follows

%%%%%%%%%%%%%%%%%%%%%%%%%%%%%%%%%%%%%%%%%%%%%%%%%%%%%
\begin{equation}
\label{Eq.nr}
\left\{ \begin{array}{l}
\displaystyle \int ^{R}_{r_0}4\pi r^2 n_+(r)dr=N_+ \\ \\
\displaystyle \int ^{R}_{r_0}4\pi r^2 n_-(r)dr=N_-,
\end{array}\right.
\end{equation}
%%%%%%%%%%%%%%%%%%%%%%%%%%%%%%%%%%%%%%%%%%%%%%%%%%%%%
%
where $r$ is the distance separation from the macroion center,
+(-) stands for counterion (coion) species and $N_{+} $ ($N_{-} $)
is the total number of counterions (coions) contained in the simulation
cell.

%%%%%%%%%%%%%%%%%%%%%%%%%%%%%%%%%%%%%
%TABLE 4
\begin{table}[b]
\caption{System parameters.}
\label{tab.runs}
\begin{ruledtabular}
\begin{tabular}{lcccccccccc}
System&
$A$&
$B$&
$C$&
$D$&
$E$&
$F$&
$G$&
$H$&
$I$&
$J$\\
\hline
$Z_m $&
60&
60&
60&
60&
60&
60&
60&
60&
180&
180\\
$Z $&
1&
1&
2&
2&
3&
3&
2&
2&
2&
2\\
$N_{+} $&
60&
60&
30&
30&
20&
20&
430&
430&
445&
445\\
$N_{-} $&
-&
-&
-&
-&
-&
-&
400&
400&
400&
400\\
\textit{$\varepsilon_2 $}&
2&
80&
2&
80&
2&
80&
2&
80&
2&
80\\
$\Delta \varepsilon  $&
78&
0&
78&
0&
78&
0&
78&
0&
78&
0\\
$R/\sigma$&
40&
40&
40&
40&
40&
40&
20&
20&
20&
20\\
\end{tabular}
\end{ruledtabular}
\end{table}
%%%%%%%%%%%%%%%%%%%%%%%%%%%%%%%%%%%%%%%%%%

Another quantity of special interest is the integrated (or cumulative)
fluid net charge $Q(r)$ defined as

%%%%%%%%%%%%%%%%%%%%%%%%%%%%%%%%%%%%%%%%%%%%%%%%%%%%%
\begin{equation}
\label{Eq.Qr}
Q(r)=\int ^{r}_{r_0}4\pi u^{2}Z\left[ n_{+}(u)-n_{-}(u)\right] du,
\end{equation}
%%%%%%%%%%%%%%%%%%%%%%%%%%%%%%%%%%%%%%%%%%%%%%%%%%%%%
%
where we chose $e=1$. $Q(r)$ corresponds to the total fluid
charge (omitting the macroion bare charge $Z_m$) within a distance
$r$ from the macroion center, and at the cell wall $Q(r=R)=Z_m$.
Up to a factor proportional to $1/r^{2}$, $\left[ Q(r)-Z_m\right]$
gives (by simple application of the Gauss theorem) the mean electric
field at $r$.
Therefore $Q(r)$ can measure the strength of the macroion
charge screening by salt-ions.
In salt-free environment systems we have
$n_-(r)=0$ and $N_{+}=Z_m/Z$.

The simulation run parameters can be found in Table
\ref{tab.runs}. For all these simulation systems, the ion
densities $n_i(r)$ were computed  with the same radial resolution
$\Delta r$. \cite{Note_Qr_resolution} The discretization of the
radial distance $r$ in $n_i(r)$  is realized over logarithmically
equidistant points so that close to the macroion surface
($r-r_0<\sigma$) we have $\Delta r<0.04\sigma$. It is important to
obtain such an accuracy (and the required statistics) if one wants
to describe quantitatively the effects of image forces which are
short-ranged at strong curvature.

\subsection{Salt-free environment}

Salt-free systems $A-F$ (see Table \ref{tab.runs}) were
investigated for a moderately charged macroion $Z_m=60$ corresponding
to a surface charge density $\sigma_0=0.11 ~ \mathrm{Cm}^{-2}$.

\subsubsection{Monovalent counterions\label{Sec.MC-monovalent}}

\begin{figure}[t]
\includegraphics[width = 8.0 cm]{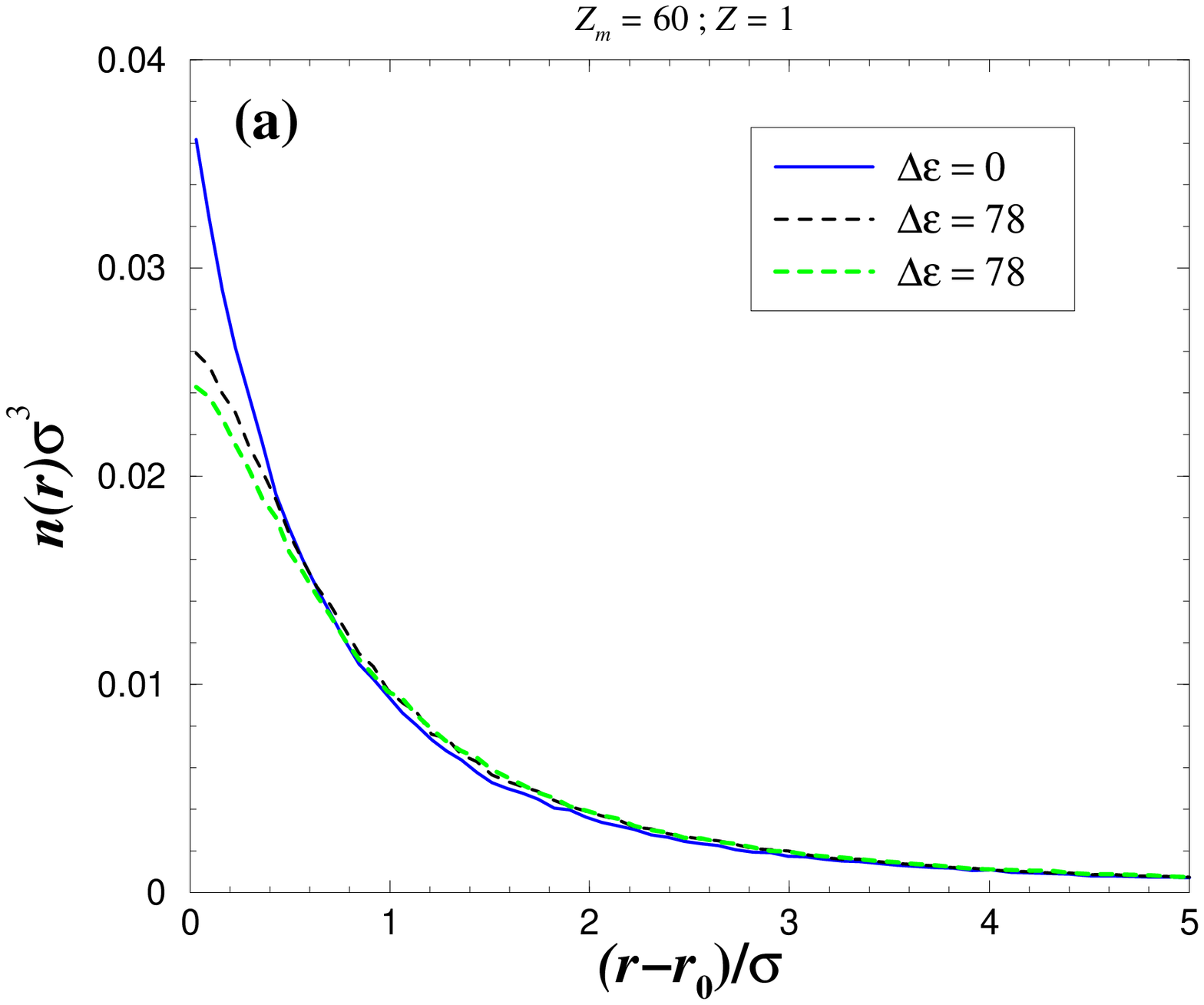}
\includegraphics[width = 8.0 cm]{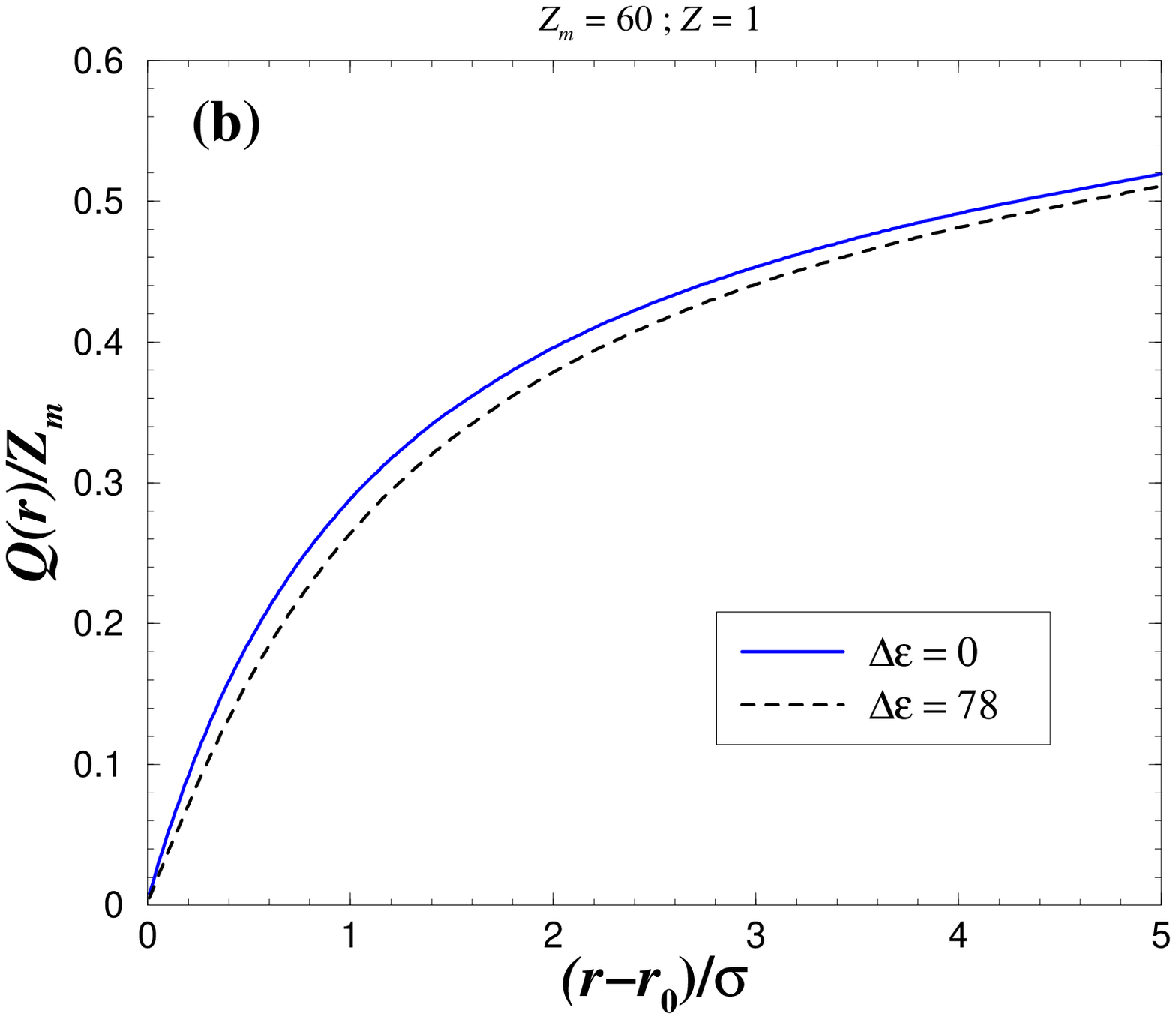}
\caption { Monovalent counterion distributions (systems $A$ and $B$):
(a) density $n_+(r)$.
The dashed line in grey corresponds to the counterion density
$n_+^{(self)}(r)$ obtained in the same system $A$
($\Delta \varepsilon = 78$) but where the (lateral) image-counterion
correlational term $U^{(im)}_{ij}$ [Eq. (\ref{Eq.MC-Uim})] has been omitted
in the total Hamiltonian $U_{tot}$ [Eq. (\ref{Eq.MC-U-tot})].
(b) fluid charge. }
\label{fig.Nr-monovalent}
\end{figure}

The profiles of $n_+(r)$ and $Q(r)$ are depicted in
Fig. \ref{fig.Nr-monovalent}(a) and (b), respectively
for the monovalent counterion systems $A$ and $B$.

Figure \ref{fig.Nr-monovalent}(a) shows that the counterion
density at contact ($r=r_0$) is somewhat  smaller  with
$\Delta\varepsilon=78$ as a direct consequence of the self-image
repulsion. However there is no maximum appearing in $n_{+}(r)$
with $\Delta\varepsilon=78$, in agreement with the study of the
single-counterion system (see Fig. \ref{fig.Vm}  and Table
\ref{tab.r*}). For $r-r_0>\sim 0.6\sigma$ (corresponding roughly
to three half ionic sizes from the interface), the effects of
image forces are negligible and all $n_{+}(r)$ curves are nearly
identical.

%%%%%%%%%%%%%%%%%%%%%%%%%%%%%%%%%%%%%%%%
%FIG 7
%%%%%%%%%%%%%%%%%%%%%%%%%%%%%%%%%%%%%%%%

To gain further insight into the effects of \textit{lateral} image-counterion
correlations, we have considered the same system $A$ ($\Delta \varepsilon=78$)
but omitted the correlational term $U^{(im)}_{ij}$
[Eq. (\ref{Eq.MC-Uim})] in the total  Hamiltonian $U_{tot}$ [Eq. (\ref{Eq.MC-U-tot})].
Physically, this means that, on the level of the image force, each counterion sees
uniquely its self-image interaction.
Thereby, Fig. \ref{fig.Nr-monovalent}(a) shows that (i) the corresponding counterion density
$n_+^{(self)}(r)$ is nearly identical to $n_+(r)$, and (ii) in the vicinity of the
interface $n_+^{(self)}(r)$ is slightly smaller than $n_+(r)$.
These findings (i) and (ii) lead to the two important conclusions:

\begin{itemize}
\item  For monovalent counterions  and moderately charged macroions,
the  \textit{effective} image force is basically identical to that of
the self-image interaction.
\cite{Note_Linse_c}
\item The crucial effect of lateral image-counterion correlations
is to \textit{screen} the self-image repulsion.
\end{itemize}
This latter feature is generally true for any \textit{finite} curvature
at identical fixed macroion charge density.
%In the limit of low curvature the trend is inversed where the
%lateral image-counterion correlations induce a stronger repulsive effective
%image force compared to that stemming from the self-image interaction alone.
%
Finding (i) is also consistent with the
fact that, close to the interface (say $r-r_0<0.2\sigma$),
the average number of (surface) counterions $\overline{N}$ is
(very) small ($\overline{N}<5$) as can be deduced from the fraction of counterions
$Q(r)/Z_m$ [Fig. \ref{fig.Nr-monovalent}(b)].

Figure \ref{fig.Nr-monovalent}(b) shows that the fluid charge
$Q(r)$ decreases when image forces are present, meaning that they
lower the macroion charge screening by counterions. At the
distance $r-r_0=\sigma$ (corresponding to a $2\sigma$-layer
thickness), the macroion is  29\% electrically compensated [i.e.,
$Q(r-r_0=\sigma )/Z_m= 0.29$] with $\Delta \varepsilon =0$ against
26\%  with $\Delta \varepsilon =78$. At the distance
$r-r_0=4\sigma$, the relative difference $\Delta Q/Q$ between the
$Q(r)$ obtained with $\Delta \varepsilon =0$ and $\Delta
\varepsilon =78$ drops to 2\% (against 10\% at $r-r_0=\sigma$)
where the bare macroion charge is nearly half-compensated.

\subsubsection{Multivalent counterions\label{Sec.MC-multivalent}}

\paragraph{Divalent counterions}

\begin{figure}
\includegraphics[width = 8.0 cm]{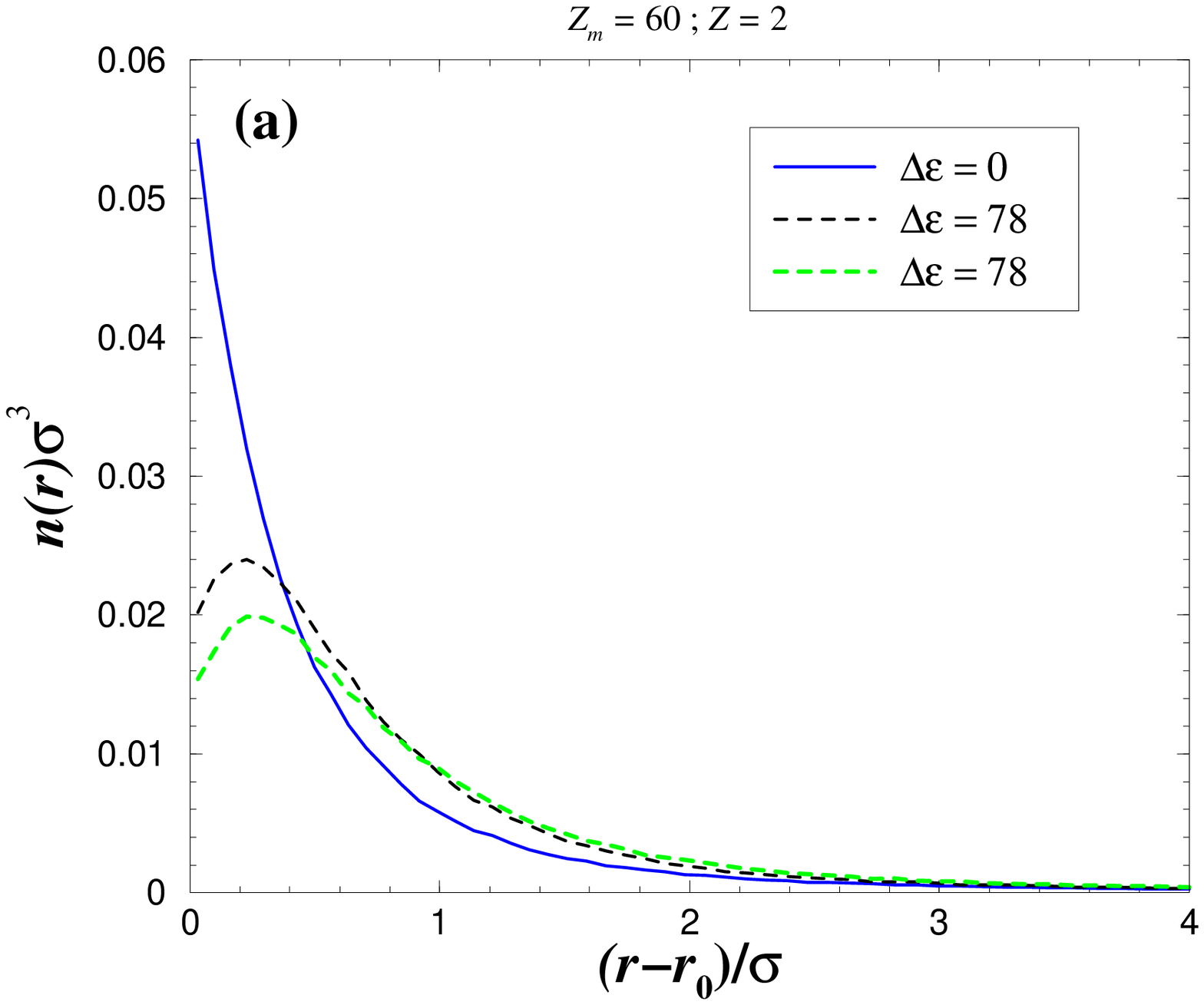}
\includegraphics[width = 8.0 cm]{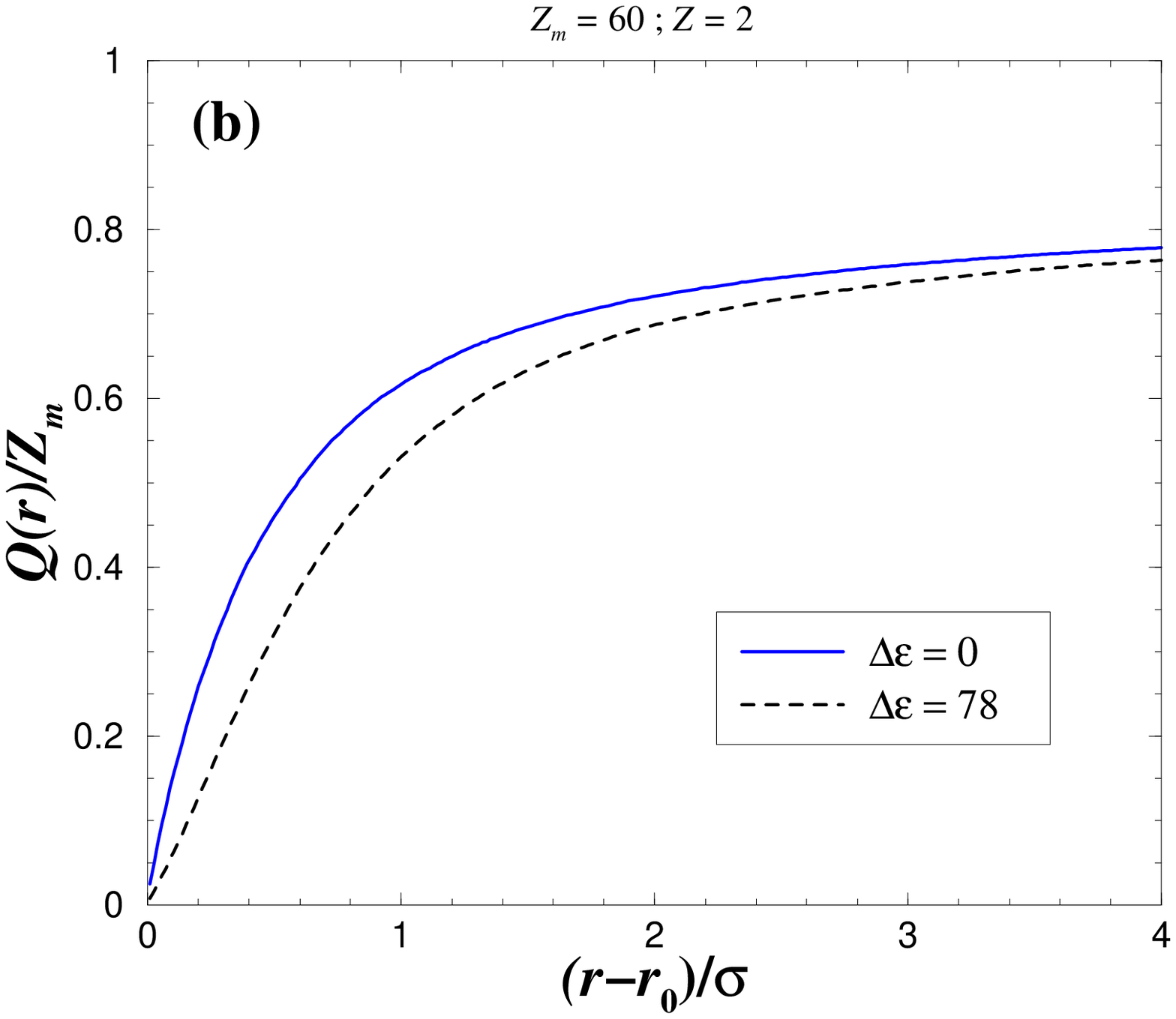}
\caption{ Divalent counterion distributions (systems $C$ and $D$):
(a) density $n_+(r)$.
The dashed line in grey corresponds to the counterion density
$n_+^{(self)}(r)$ obtained in the same system $C$
($\Delta \varepsilon = 78$) but where the (lateral) image-counterion
correlational term $U^{(im)}_{ij}$ [Eq. (\ref{Eq.MC-Uim})] has been
omitted in the total Hamiltonian $U_{tot}$ [Eq. (\ref{Eq.MC-U-tot})].
(b) fluid charge. }
\label{fig.Nr-divalent}
\end{figure}

The profiles of $n_+(r)$ and $Q(r)$ are depicted in Fig.
\ref{fig.Nr-divalent}(a) and (b), respectively for the divalent
counterion systems $C$ and $D$.

%%%%%%%%%%%%%%%%%%%%%%%%%%%%%%%%%%%%%%%%%%%%%%%%%
%FIG 8
%%%%%%%%%%%%%%%%%%%%%%%%%%%%%%%%%%%%%%%%%%%%%%%%%

Figure \ref{fig.Nr-divalent}(a) shows that the counterion density
at contact becomes strongly reduced with $\Delta \varepsilon=78$
due to the $Z^2$-dependence of the self-image repulsion
[compare the case  $Z=1$ in Fig. \ref{fig.Nr-monovalent}(a)].
This sufficiently strong (short-ranged) repulsion leads to a maximum in $n_+(r)$
close to the macroion surface.
The corresponding radial position $r^*$ maximizing $n_+(r)$ is $r^*=r_0+0.22\sigma$,
in excellent agreement (within $\Delta r$) with the
\textit{one-counterion} theoretical value $r_0+0.17\sigma$ (see Table \ref{tab.r*}).
This shows that for divalent
counterions many-body effects do nearly not affect $r^*$.
This non-trivial finding is the result of the competition
between two driving forces that control $r^*$ in
\textit{many}-counterion systems:
\begin{itemize}
\item $F_{im}$: the screening of the self-image \textit{repulsion}
by the (extra) negative polarization charges  tends to \textit{decrease}
the $r^*$ obtained in the one-counterion system.

\item $F_{mc}$: the screening of the macroion-counterion \textit{attraction} by
the (extra) surface counterions tends to \textit{increase}
the $r^*$ obtained in the one-counterion system.
\end{itemize}
It is precisely a balance of these two driving forces
that leads to a nearly unchanged $r^*$ (compared to the one-counterion system)
in many-counterion systems.
Whereas for monovalent counterions both driving forces
$F_{im}$ and $F_{mc}$ are weak, those become relevant for multivalent counterions.
%since their strength scales like $Z^2$.

We stress the fact that this is specific to the spherical geometry,
and that for a planar interface (at identical surface charge density)
one should get a higher $r^{*}$ (compared to that of the one-counterion system),
since there we have no screening driving force $F_{im}$.
We are not aware of any previous studies for the planar interface
that address this issue. \cite{Note_Torrie}

To gain even further insight into the effect of $Z$ on the lateral
image-counterion correlations, we have ignored the term $U^{(im)}_{ij}$
in $U_{tot}$ in the same system $D$ ($\Delta \varepsilon = 78$) as done previously
with system $A$.
Figure \ref{fig.Nr-divalent}(a) shows a qualitatively different $n_+^{(self)}(r)$
where $r^{*}=r_0+0.26\sigma$ is now somewhat larger,
proving that with divalent counterions the screening of the self-image
repulsion by lateral image-counterion correlations is appreciable.
This is in contrast to what was observed with $Z=1$.

At the distance $r-r_0=\sigma$, Fig. \ref{fig.Nr-divalent}(b) shows that
the macroion is 62\% electrically compensated for
$\Delta \varepsilon =0$ against 53\% for $\Delta \varepsilon =78$
[compare the case $Z=1$ in Fig. \ref{fig.Nr-monovalent}(b)].

\paragraph{Trivalent counterions}

\begin{figure}
\includegraphics[width = 8.0 cm]{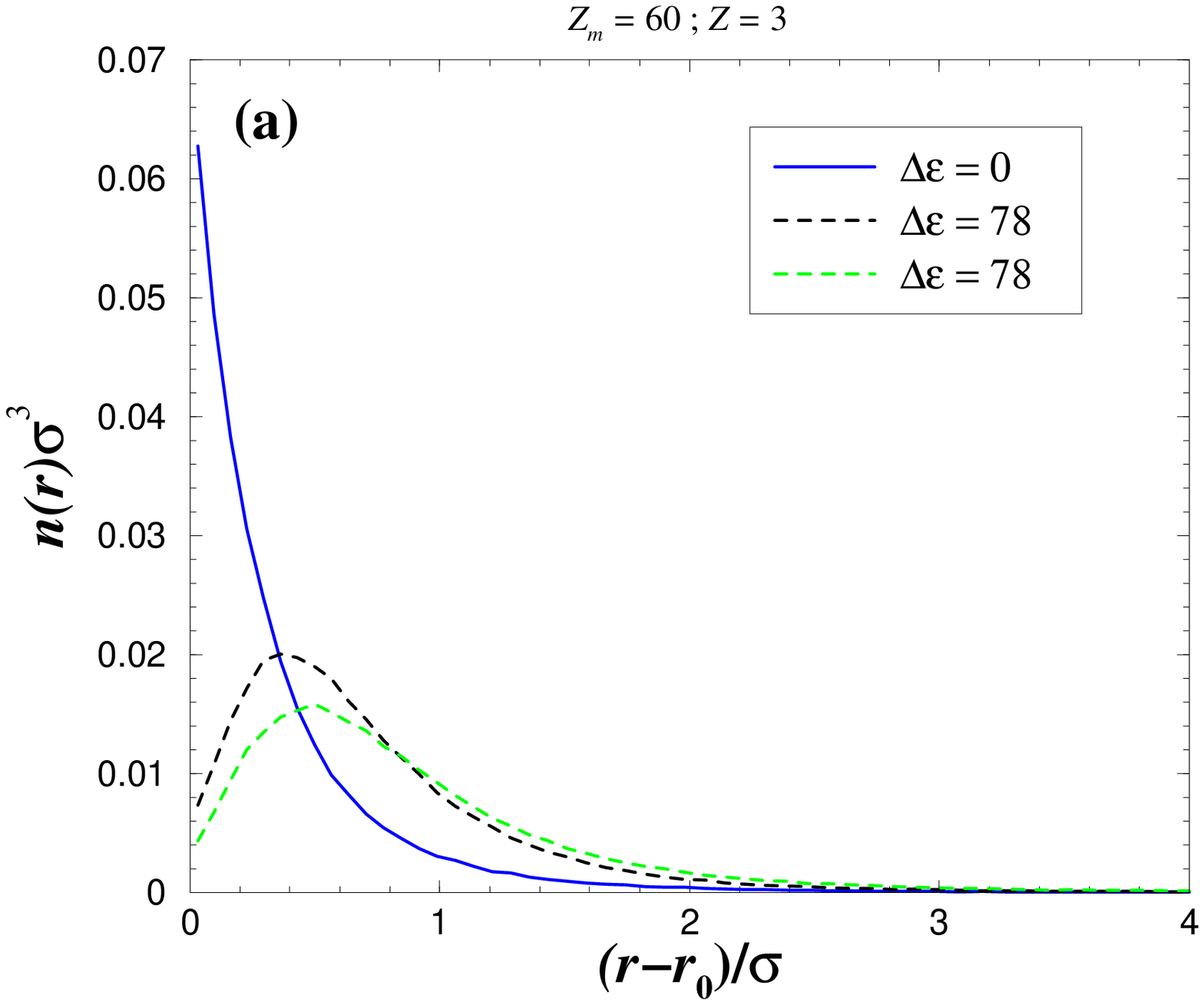}
\includegraphics[width = 8.0 cm]{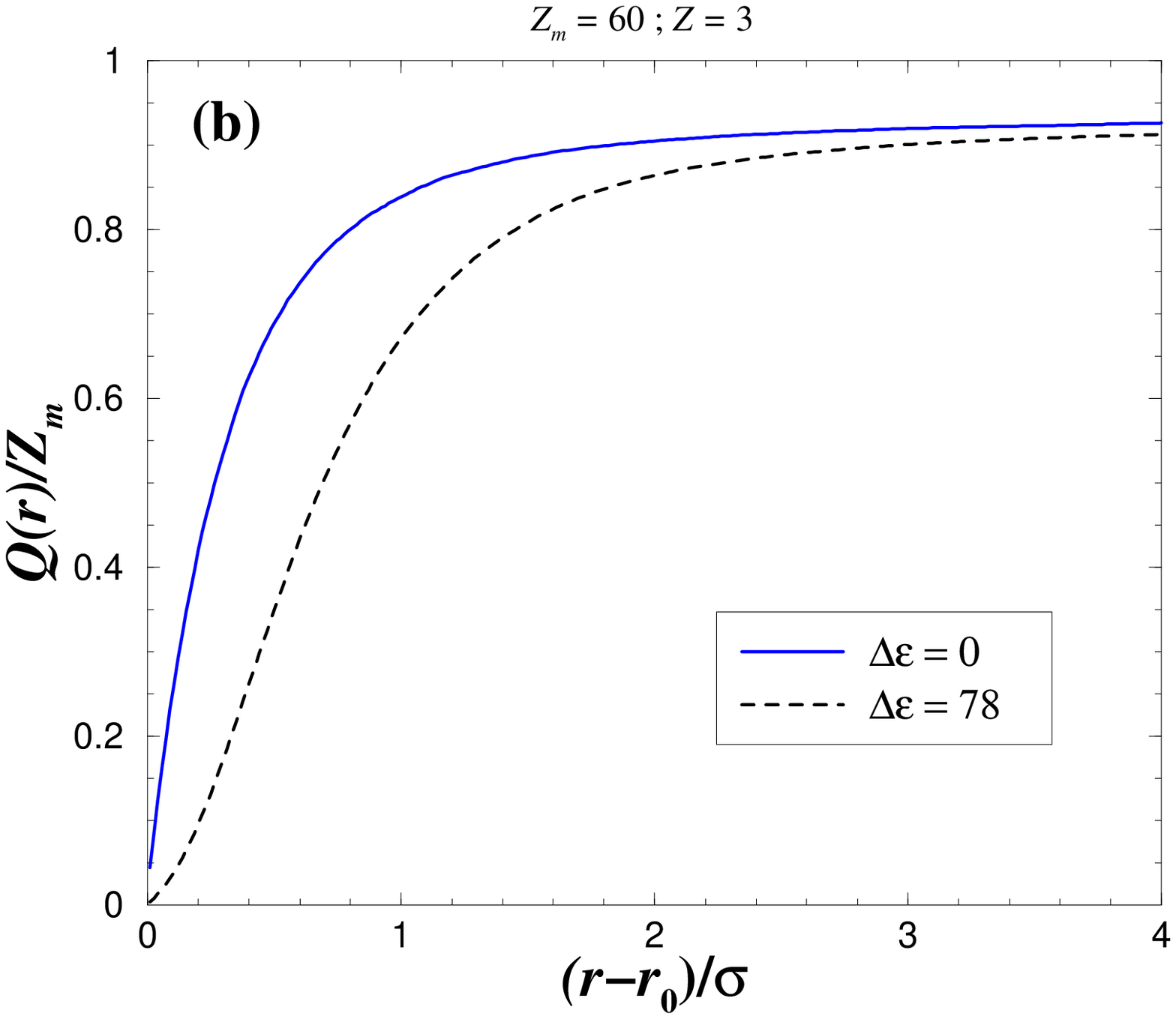}
\caption{ Trivalent counterion distribution (systems $E$ and $F$):
(a) density $n_+(r)$.
The dashed line in grey corresponds to the counterion density
$n_+^{(self)}(r)$ obtained in the same system $E$
($\Delta \varepsilon = 78$) but where the (lateral) image-counterion
correlational term $U^{(im)}_{ij}$ [Eq. (\ref{Eq.MC-Uim})] has been omitted
in the total Hamiltonian $U_{tot}$ [Eq. (\ref{Eq.MC-U-tot})].
(b) fluid charge. }
\label{fig.Nr-trivalent}
\end{figure}

The profiles of $n_+(r)$ and $Q(r)$ are depicted in
Fig. \ref{fig.Nr-trivalent}(a) and (b), respectively
for trivalent counterion systems $E$ and $F$.

%%%%%%%%%%%%%%%%%%%%%%%%%%%%%%%%%%%%%%%%%%%%%%%
%FIG  9
%%%%%%%%%%%%%%%%%%%%%%%%%%%%%%%%%%%%

Figure \ref{fig.Nr-trivalent}(a) shows that the counterion density
at contact is drastically reduced with  $\Delta \varepsilon=78$,
as expected for high $Z$ (compare the previous cases).
At $\Delta \varepsilon =78 $, we have
$r^{*}=r_0+0.36\sigma$, in quantitative agreement with the \textit{one-counterion}
theoretical value $r_0+0.32\sigma$ (see Table \ref{tab.r*}).
This shows again that even for trivalent counterions many-body effects
do (practically) not affect $r^{*}$ (compared to that obtained in the single-counterion system)
due to a balance of the driving forces $F_{im}$ and $F_{mc}$.
%The same behavior is expected for higher $Z $ since the screening
%of the positive polarization charges increases with $Z $.

By neglecting the lateral image-counterion correlations
in the same system  $E$ ($\Delta \varepsilon=78$),
Fig. \ref{fig.Nr-trivalent}(a) indicates that the position $r^*$ of the
maximum in $n_+^{(self)}(r)$ gets considerably larger ($r^*=r_0+0.50\sigma$).
This relatively strong shift confirms the $Z$-enhancing of the  screening
of the self-image repulsion by lateral image-counterion correlations.

\begin{figure}
\includegraphics[width = 7.0 cm]{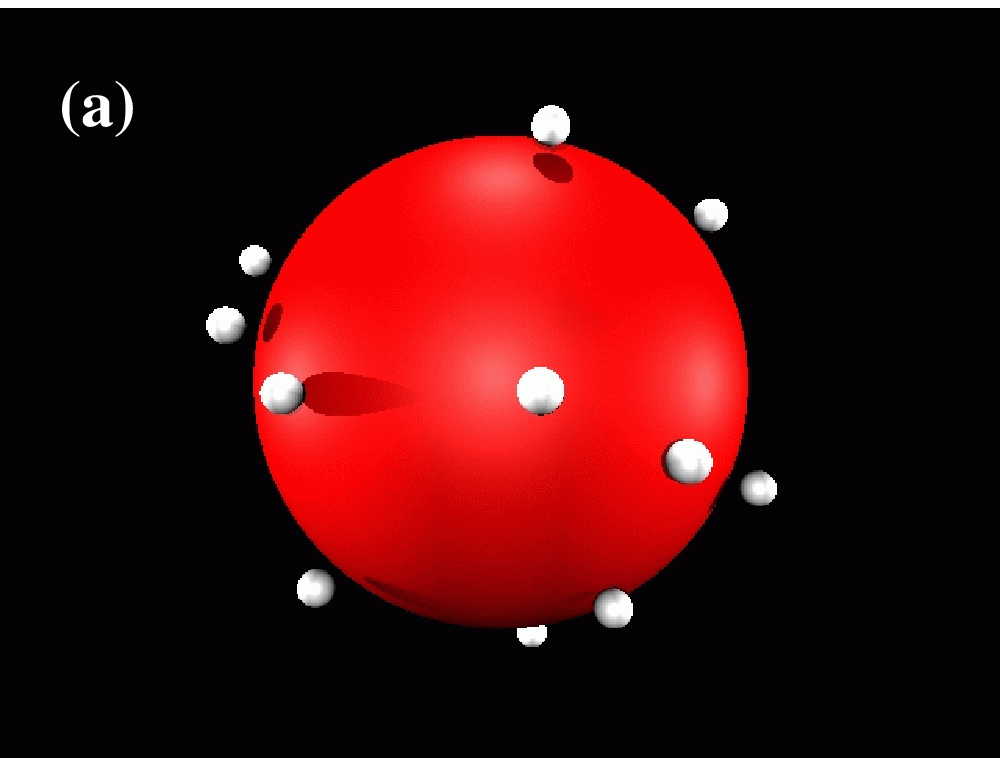}
\includegraphics[width = 7.0 cm]{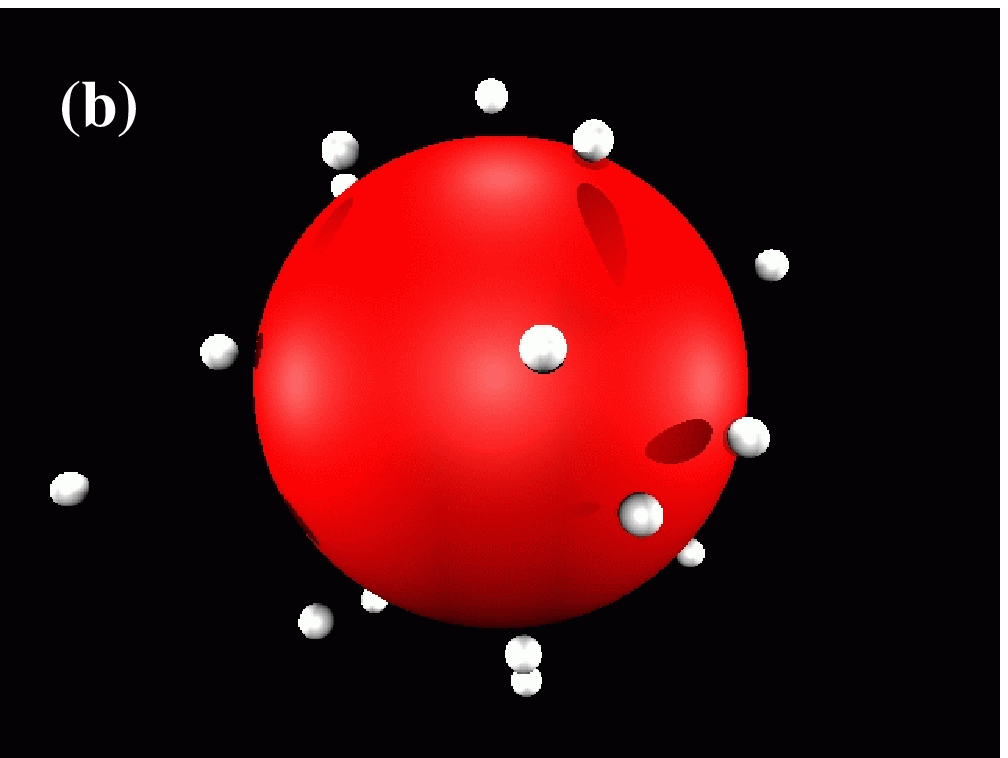}
\caption{
Snapshots of typical equilibrium configurations for trivalent counterions
(systems $E$ and $F$).
(a) $\Delta \varepsilon =0$
(b) $\Delta \varepsilon =78$.
One can clearly observe
the larger mean radial counterion distance for $\Delta \varepsilon =78$
stemming from the self-image repulsion.
%Note also the large \textit{lateral}
%separation between counterions due to the strong electrostatic repulsion
%that occurs with $ Z=3 $.
}
\label{fig.snap-trivalent}
\end{figure}

At the distance $r-r_0=\sigma$, the macroion is  84\% electrically
compensated for $\Delta\varepsilon=0$ against only 67\% for
$\Delta\varepsilon=78$ [see Fig. \ref{fig.Nr-trivalent}(b) and
compare previous systems]. Snapshots of typical equilibrium
configurations for $\Delta\varepsilon=0$ and
$\Delta\varepsilon=78$ can be visualized in Fig.
\ref{fig.snap-trivalent} (a) and (b), respectively.

%%%%%%%%%%%%%%%%%%%%%%%%%%%%%%%%%%%%
%FIG 10
%%%%%%%%%%%%%%%%%%%%%%%%%%%%%%%%%%%%%%%%%%%%%%%

\subsection{Salty solutions\label{Sec.MC-salt}}

We  focus on the case of divalent salt-ions.
This choice is motivated by two reasons:
(i) effects of image charges are clearly observable for multivalent
counterions and
(ii) such systems must be experimentally reachable.
To study the effect of added salt we have considered  two macroion charges
$Z_m=60$ (as previously) and $Z_m=180$ corresponding to a charge density
$\sigma_0=0.32 ~ \mathrm{Cm^{-2}}$.
The salt concentration defined as $\frac{N_-}{\frac{4}{3} \pi R^3}$ is
$0.44$ M for all salty systems $G-J$
(see Table \ref{tab.runs}).
The simulation cell radius $R=20\sigma$ of these systems
is still very large compared to any screening lengths so that finite size effects
are negligible.

\subsubsection{Moderately charged macroion }

\begin{figure}
\includegraphics[width = 8.0 cm]{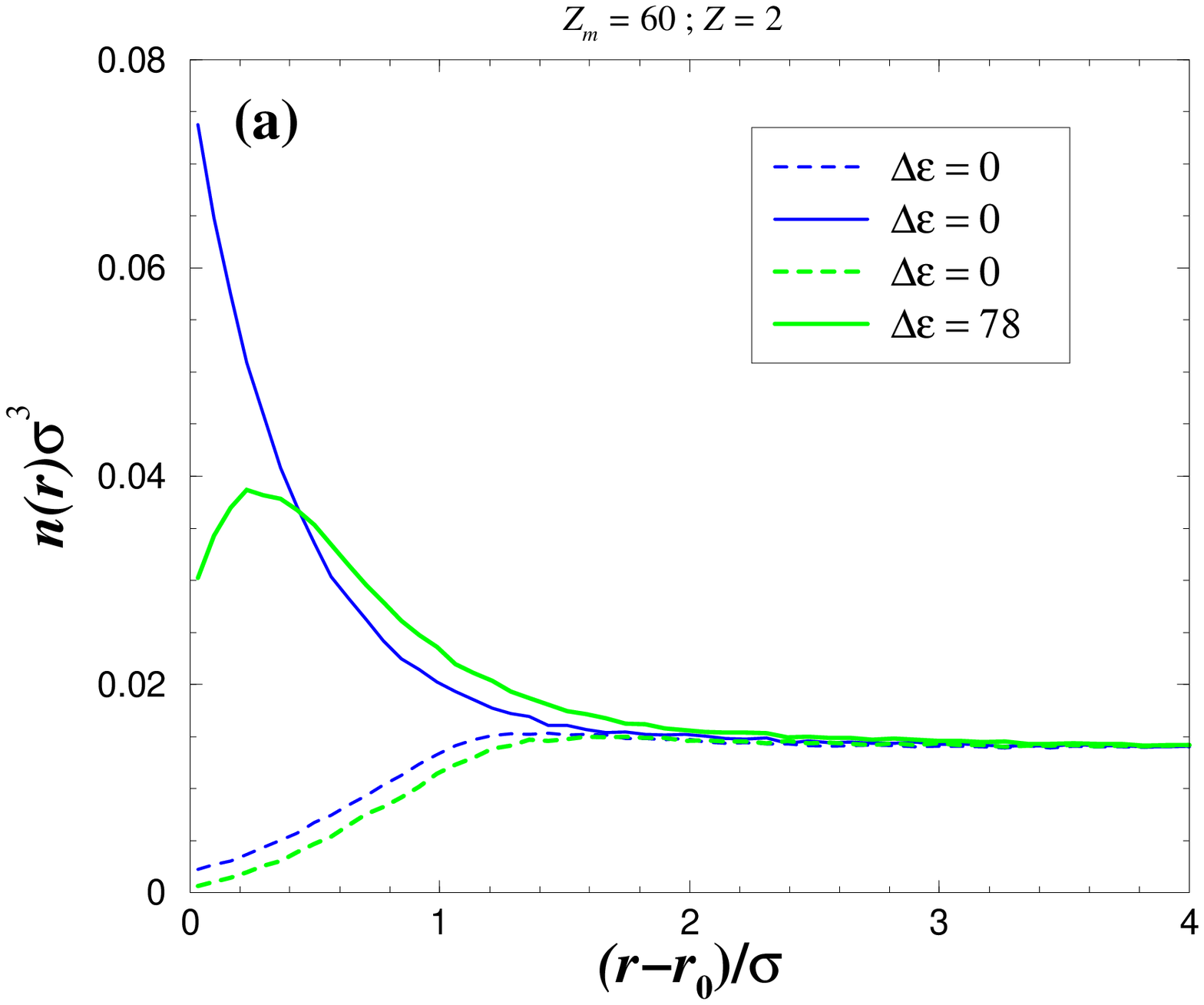}
\includegraphics[width = 8.0 cm]{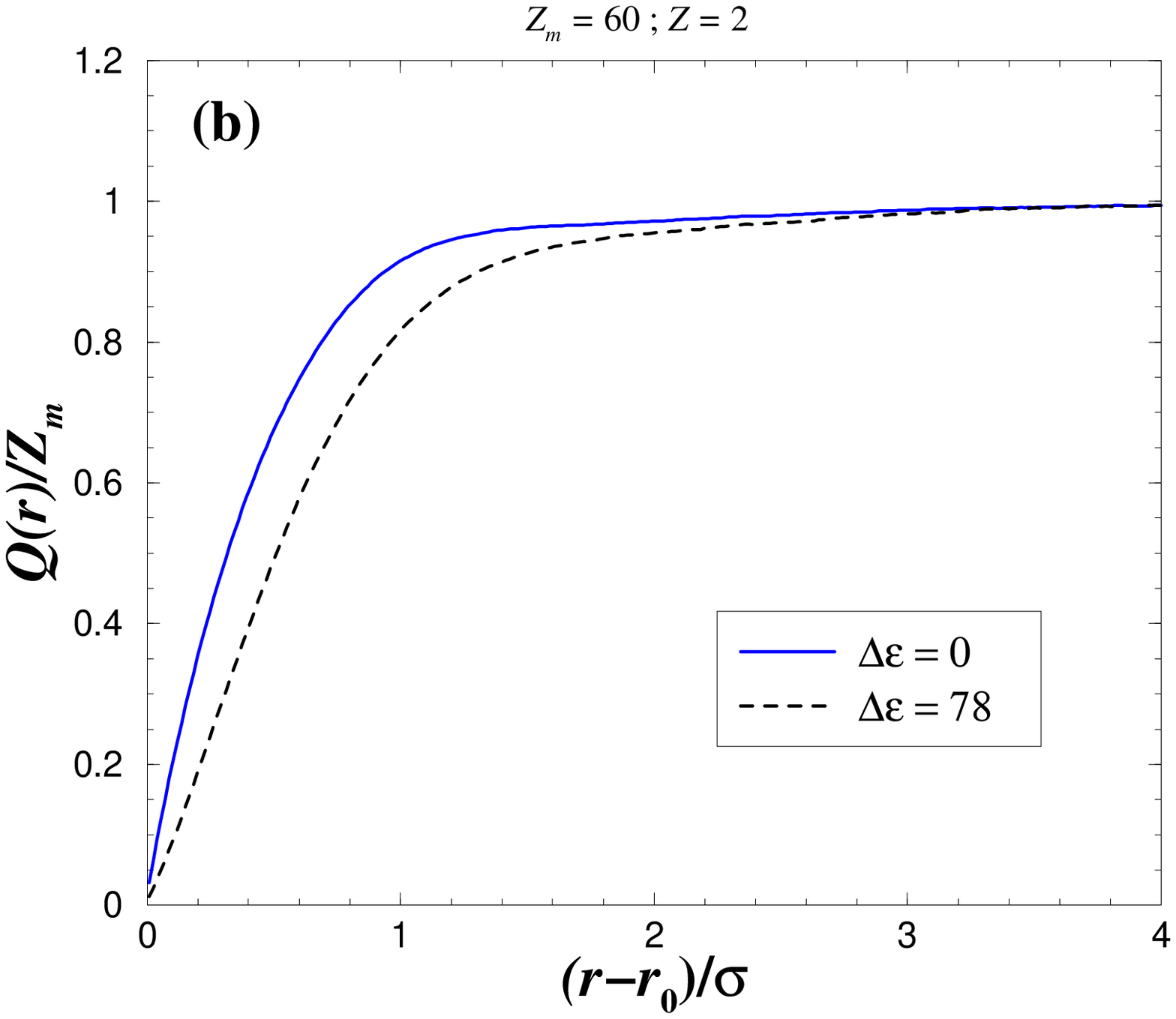}
\caption{
Divalent salt-ion distribution (systems $G$ and $H$) with $Z_m=60$:
(a) The solid and dashed lines correspond to counterion and coion
densities, respectively.
(b) net fluid charge.
}
\label{fig.Nr-Z60-salt}
\end{figure}

Profiles of $n_{\pm}(r)$ and $Q(r)$ are depicted in
Fig. \ref{fig.Nr-Z60-salt}(a) and (b), respectively
for the salty systems $G$ and $H$ with $Z_m=60$.

The coion density $n_-(r)$ with $\Delta\varepsilon=78$ is
basically shifted to the right of about $0.15\sigma$ (compared to
that with $\Delta\varepsilon=0$) due to the repulsive coion'
self-image interaction. Near  the colloidal surface, the
counterion densities $n_{+}(r)$ are considerably higher than
those obtained with no added salt (systems $C$ and $D$) as it
should be [compare Fig. \ref{fig.Nr-divalent}(a)].

A rather surprising result here is that, despite of the presence of a considerable
amount of added salt, we still have $r^*=r_0+0.22\sigma$  remaining unchanged.
%that obtained in the one-counterion system ($r_0+0.17\sigma$).
%
This is a non-trivial finding since one should have an (extra) \textit{attractive}
contribution to the macroion-counterion potential of mean force stemming from
the (localized) \textit{negative} polarization charges
induced by the coions, which in turn could lead to a shorter $r^*$.
However there are two concomitant sources that lead to a marginal
screening of the counterion' self-image repulsion
by the negative coion-induced polarization charges:
(i) there is a strong coion depletion close to
the interface [see Fig. \ref{fig.Nr-Z60-salt}(a)] due to the
large direct Coulomb macroion-coion repulsion and
(ii) $|\sigma_{pol}^{(sph)}|$ decreases abruptly with the radial distance
of the microion as discussed in Sec. \ref{Sec.induced-charge}
(see also Fig. \ref{fig.Qpol}).
Of course the role of the \textit{excluded volume} is crucial here.

As expected the macroion charge screening is weaker when  image forces
come into play as can be deduced from the profile of $Q(r)$ plotted
in Fig. \ref{fig.Nr-Z60-salt}(b).

%%%%%%%%%%%%%%%%%%%%%%%%%%%%%%%%%%%%%%%%%%%
%FIG 11
%%%%%%%%%%%%%%%%%%%%%%%%%%%%%%%%%%%%%%%%%%%

\subsubsection{Highly charged macroion}

\begin{figure}
\includegraphics[width = 8.0 cm]{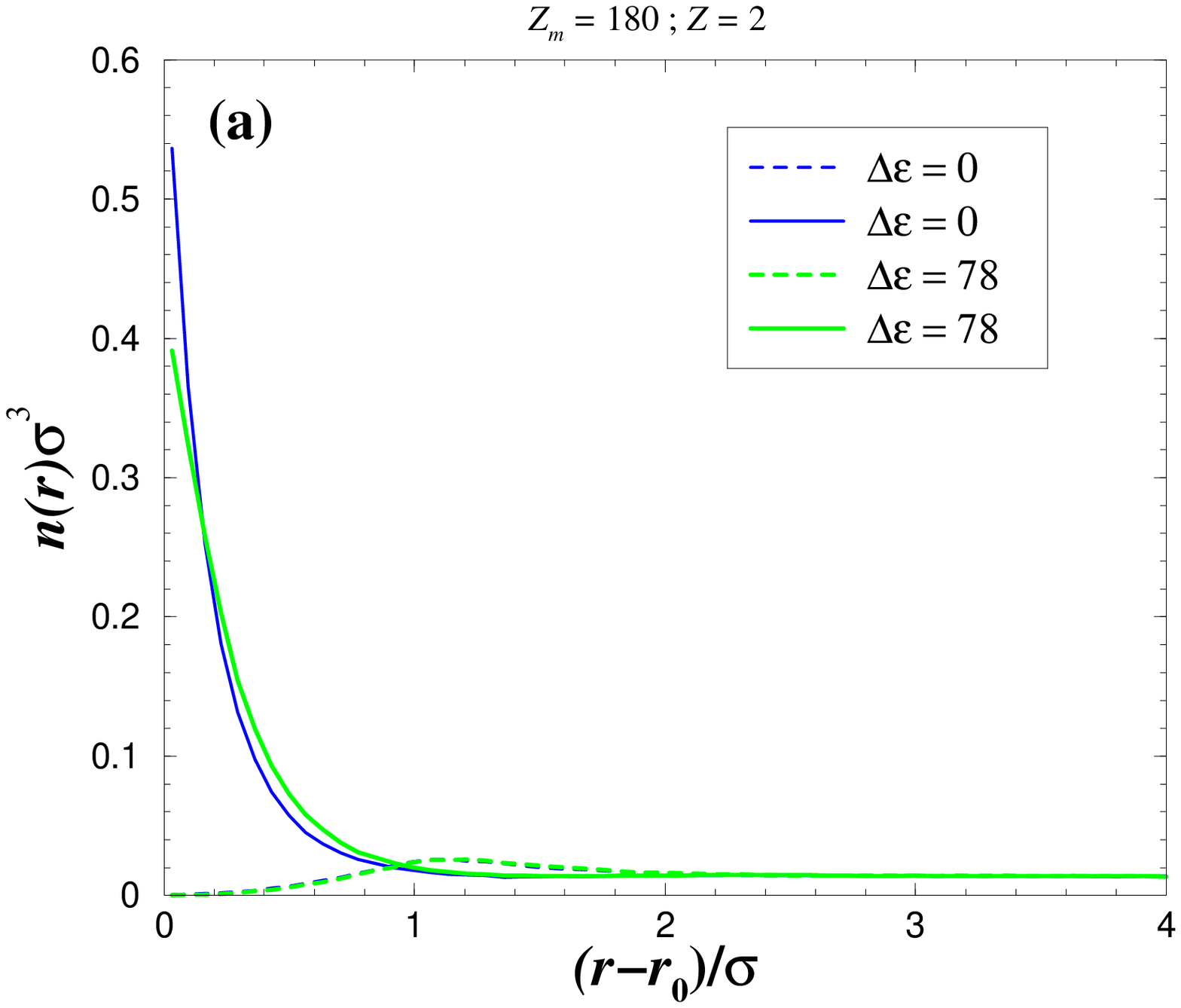}
\includegraphics[width = 8.0 cm]{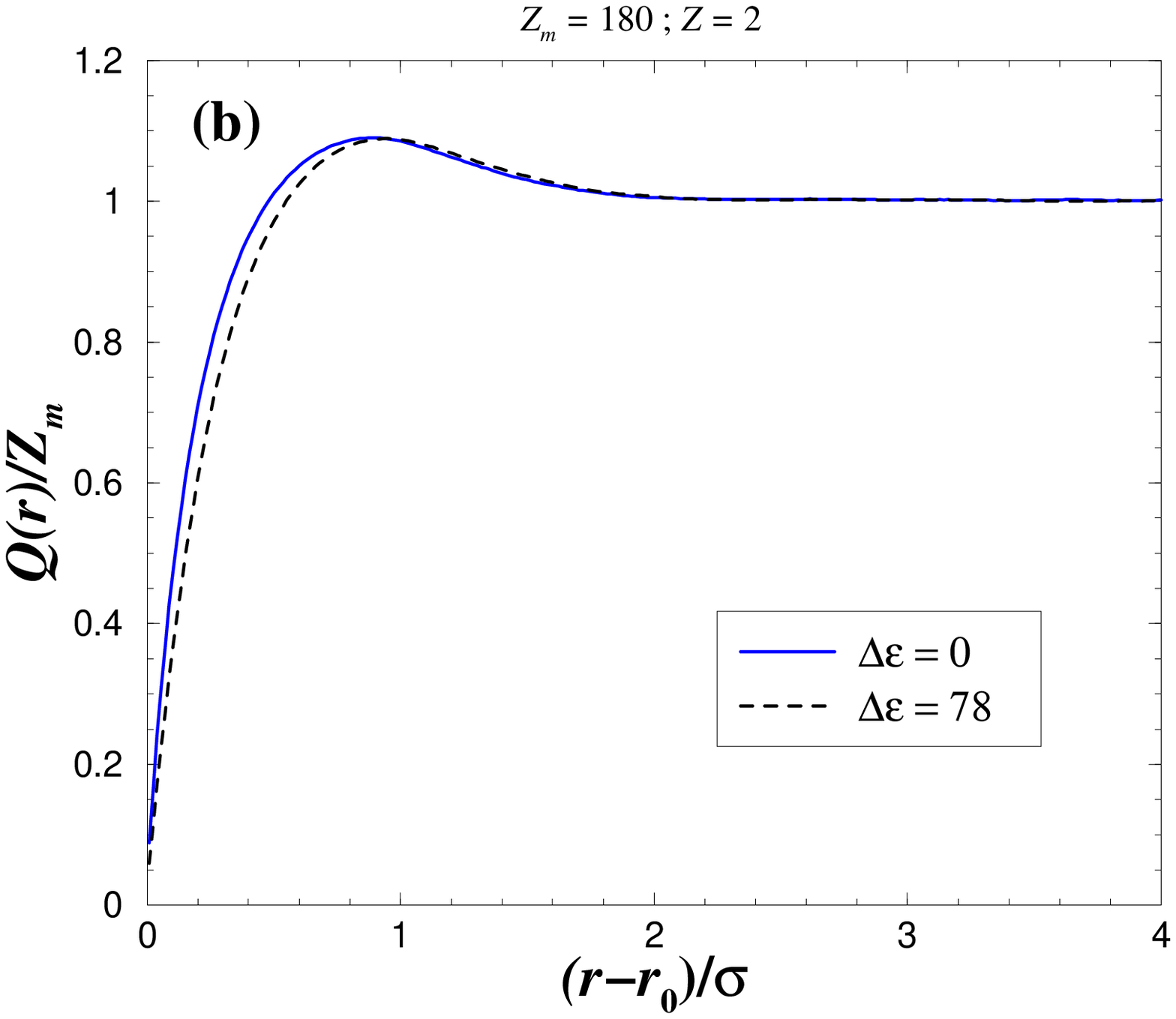}
\caption{
Divalent salt-ion distribution (systems $I$ and $J$) with $Z_m=180$:
(a) The solid and dashed
lines correspond to counterion and coion densities, respectively.
(b) net fluid charge.
}
\label{fig.Nr-Z180-salt}
\end{figure}

Profiles of $n_{\pm}(r)$ and $Q(r)$ are depicted in
Fig. \ref{fig.Nr-Z180-salt}(a) and (b), respectively
for the salty systems $I$ and $J$ with $Z_m=180$.

%%%%%%%%%%%%%%%%%%%%%%%%%%%%%%%%%%%%%%%%%%%%
% FIG 12
%%%%%%%%%%%%%%%%%%%%%%%%%%%%%%%%%%%%%%%%%%%%%%

Figure \ref{fig.Nr-Z180-salt}(a) shows that the effects of image forces
are considerably reduced.
The relatively small difference between the $n_+(r)$ obtained with $\Delta\varepsilon=0$
and that obtained with $\Delta\varepsilon=78$ decreases drastically
in the vicinity of the interface, and already for
$r-r_0>\sim 0.2\sigma$ the two profiles of $n_{+}(r)$ are nearly identical.
Besides, near the interface \textit{no} effective
macroion-counterion repulsion occurs  at $\Delta\varepsilon=78$.
This absence of a maximum in $n_+(r)$ is due to two main concomitant effects:

% 2 MECHANISMS
\begin{itemize}
\item For such a highly charged macroion, there is a very large number
of counterions close to the interface
[compare Fig. \ref{fig.Nr-Z180-salt}(b) and Fig. \ref{fig.Nr-Z60-salt}(b)].
In this limit, one can use Wigner crystal concepts and say that, on the
level of the force stemming from the bare charges
(i.e, ignoring the image forces),
each surface counterion essentially interacts with the oppositely charged
background of its Wigner-Seitz (WS) cell.
At sufficiently high macroion charge density (i.e., small WS hole radius),
this attractive interaction becomes very important and it always overcomes
the self-image repulsion.
%This is precisely our first mechanism that
%accounts for the absence of a maximum in $n_+(r)$.
%
\item The second (concomitant) mechanism is specific to the closed spherical topology:
at high number of surface counterions, the  image forces
are reduced because of the enhanced degree of spherical symmetry
as already mentioned in Sec. \ref{Sec.Vm}.
\end{itemize}
%
%
%This latter effect becomes especially relevant at large colloidal curvature as
%it is presently the case.

%The role of excluded volume is again crucial here since it
%allows a finite value for the potential of self-iùmage interaction.
%The same mechanism applies to planar geometry where it was also found that,
%for highly charged plates, no maximum  appears in the counterion density
%\cite{Torrie_JCP_1982}.

%at high number of surface counterions, the inhomogeneity as well as the magnitude
%of the (total) counterion-induced polarization charge becomes much weaker
%(compared to the one-counterion case) as already thoroughly discussed.

The coion densities $n_{-}(r)$ are basically identical for both dielectric
discontinuities $\Delta\varepsilon$, in contrast to what happened
with $Z_m=60$ (systems $J$ and $K$).
This non-trivial finding  can be explained as the enhanced screening of
the coion' self-image repulsion by the positive polarization charges
induced by the other coions present in the electrical double layer (EDL).
Indeed, because of the macroion charge \textit{reversal} that occurs at
$Z_m=180$ [i.e., $Q(r)/Z_m>1$ - see Fig. \ref{fig.Nr-Z180-salt}(b)],
there is  also a larger number of coions (at fixed salt concentration) in the EDL
[compare Fig. \ref{fig.Nr-Z180-salt}(a) and Fig. \ref{fig.Nr-Z60-salt}(a)].
Therefore, since the magnitude and the inhomogeneity of
$-\sigma_{pol}^{(sph)}(\theta)$
induced by a coion strongly decreases with its radial distance
[see Eq. (\ref{Eq.pol-charge-FULL}) and Fig. \ref{fig.Qpol}],
the screening of the coion' self-image repulsion gets highly sensitive
to an increase in number of coions in the EDL.
%total (less localized) negative coion-induced polarization
%charge by the positive one is much more sensitive to the increase in number
%of coions in the EDL.

Concerning the net fluid charge $Q(r)$, we see that  both profiles
obtained with $\Delta \varepsilon = 78$ and  $\Delta \varepsilon =
0$ are nearly identical, as expected from those of $n_{\pm}(r)$.
The net fluid charge $Q(r)$ reaches its maximum at
$r^*_{Q}-r_0=0.90\sigma$ and $0.94\sigma$ for
$\Delta\varepsilon_2=0$ and 78, respectively. In both cases we
have a macroion charge reversal of 9\% {[}more explicitly
$Q(r^*_{Q})/Z_m=1.09${]}. This proves the important result that,
for typical systems (with high macroion charge density) leading to
overcharging,
\cite{Messina_EPJE_2001,Messina_PhysicaA_2002,Messina_PRL_2000,
Messina_EPL_2000,Messina_PRE_2001} image forces do \textit{not}
affect the strength of the macroion charge reversal.
% but it only slightly shifts the radial position $r^*_{Q}$.

%%%%%%%%%%%%%%%%%%%%%%%%%%%%%%%%%%%%%%%%%%%%%%
\section{Concluding remarks \label{Sec.conclu}}
%%%%%%%%%%%%%%%%%%%%%%%%%%%%%%%%%%%%%%%%%%%%%%

We have presented fundamental results about the effects of image forces
on the counterion distribution around a spherical macroion.

Exact analytical results have been provided for the case of a
single microion interacting with a dielectric sphere. Within this
framework, the self-image interaction and the surface charge of
polarization have been studied and also compared to those obtained
with a planar interface. Besides we also estimated the position
$r^*$ where the macroion-counterion potential of interaction is
minimized. We demonstrated that the effects of image forces due to
a spherical interface are qualitatively different from those
occurring with a planar interface, especially when the colloidal
curvature is large. We showed that the  \textit{self-screening} of
the polarization charges (i.e., the screening of the positive
surface charges of polarization by the negative ones) is decisive
to explain the weaker and the shorter range of the self-image
interaction in spherical geometry. This
self-screening increases with the colloidal curvature.
%All our quantitative findings concerning a single charged (or uncharged) spherical colloid
%should also qualitatively apply, in the limit of dilute colloidal concentration and weak
%charge density, to realistic many-colloid systems neutralized by their counterions.
%The \textit{self-screening} of
%the polarization charge (i.e., screening of the positive surface polarization
%charges by the negative ones) is stronger the higher the colloidal
%curvature.

Many-counterion systems have been investigated by means of extensive MC
simulations where image forces were properly taken into account.

In salt-free environment and  for moderately charged macroions, a
maximum in the counterion density (near the spherical interface)
appears for sufficiently large dielectric discontinuity
$\Delta\varepsilon$. An important result is that the corresponding
position $r^*$ is basically identical, regardless of the
counterion valence $Z$, to that obtained within the
\textit{one}-counterion system. This feature is specific to the
spherical geometry and can not take place with planar interfaces where
there is \textit{no} self-screening of the polarization charges.
For monovalent counterions  we showed that the (effective)
image force is basically equal to that of the self-image interaction,
and the \textit{lateral} image -counterion correlations are (very)
weak. However for multivalent counterions the lateral
image-counterion correlations affect significantly the counterion
density, and as major effect they \textit{screen} the self-image repulsion.
Nevertheless, the combined effects of (i) the macroion
charge screening by counterions and (ii) the screening of the
self-image repulsion lead to a nearly unchanged $r^*$ (compared to
that obtained in the single-counterion system) for multivalent
many-counterion systems.
We also showed that the counterion density at contact decreases
drastically with $Z$, and that $r^*$ also increases with $Z$ as
expected. These latter results have important implications for the
stabilization of charged colloidal suspensions where a component
of the pair-force is proportional to the ion density at contact.

By adding salt, it was found for moderately charged macroions that
the strength of the image forces induced by the \textit{coions} is
very small compared to that resulting from the counterions. This
is due to the coupled effects of (i) the coion depletion in the
vicinity of the colloidal interface due to the strong direct
Coulomb macroion-coion repulsion and (ii) the (highly) short range
of the image forces in spherical geometry. Consequently the
position $r^*$ remains identical to that obtained in salt free
environment and a fortiori to that obtained within the
one-counterion system. For \textit{highly} charged macroions the
effects of image charges are significantly reduced since (i) the
attractive counterion-hole interaction dominates  the repulsive
counterion' self-image interaction and (ii) the screening of the
counterion' self-image repulsion gets enhanced by symmetry reason.
In this situation \textit{no} maximum  appears in the counterion
density and it was found that \textit{overcharging} is nearly
unaffected by image forces.

%The only noticeable effect is to slightly shift (from about
%1/10 atomic size) the radial profile of the net fluid charge, but
%the strength of macroion charge reversal is identical.

Although our MC analysis was carried at given macroion size, all
the above reasonings that concern \textit{many} counterions remain
unchanged (for symmetry reason) for any \textit{finite} curvature
by a rescaling at fixed macroion charge density.

Finally, this contribution should constitute a solid basis to understand
and predict the effects of image charges in other similar systems
(e.g., polyelectrolyte adsorption onto spherical charged colloids).

\acknowledgments
I thank C. Holm, K. Kremer and H. Schiessel for valuable discussions.
This work had been supported by \textit{Laboratoires Europ\'eens Associ\'es} (LEA).

%\bibliographystyle{prsty}
%\bibliography{colloid}

\begin{thebibliography}{10}
\bibitem{Note_Eq_1overR}
Equation (\ref{Eq.1overR}) is also of the form of
Eq. (\ref{Eq.FI-general}).
\bibitem{Note_zero_monopole}
Mathematically, this is a direct consequence of the orthogonality
of the Legendre polynomials. Indeed, we have $\int ^{1}_{-1}P_{l}(x)P_{0}(x)dx=0$
and since $P_{0}(x)=1$ it results from this that $\int ^{1}_{-1}P_{l}(x)dx=0$.
\bibitem{Note_convergence_pol}
It turns out that the convergence of the Legendre sum in
Eq. (\ref{Eq.pol-charge-FULL}) is extremely slow.
With $l_{max}=300$ we have a relative error of $10^{-6}$ in 
the most severe case where $b=8\sigma$.
\bibitem{Note_short_range}
This holds when the microion $q$ is sufficiently far away
from the interface so that only the dipolar term matters (i.e, $l=1$).
However close to the colloidal surface one has to take all higher multipolar
contributions to get the exact potential of interaction.
\bibitem{Note_comparison_2methods}
We have carefully checked that both methods {[}i.e., Eq. (\ref{Eq.FI-out-FULL})
and Eq. (\ref{Eq.FI-out-IMG}){]} are numerically identical. However,
at large $a$ (typically $a>30\sigma$) the Legendre sum
in Eq.~\eqref{Eq.Vr-sphere} becomes numerically unstable.
\bibitem{Note_Linse_b}
Note that the two-image charge approximation also holds when the charge
$q$ is very far from the interface (i.e., $b/a\gg 1$). Unfortunately,
all the relevant effects of image forces in spherical geometry occur
within the opposite limit $b/a\rightarrow 1$.
\bibitem{Note_plate}
In the case of a charged plate, to be consistent, one has to think either of a
large finite plate in the presence of a single microion, or equivalently of an
infinite plate in the presence of a very dilute dispersion of microions.
\bibitem{Note_convergence_Vself}
Here, the convergence of the Legendre sum in 
Eqs.~\eqref{Eq.MC-Uself} and \eqref{Eq.MC-Uim}
is much faster, in contrast to what happened with the computation of
$\sigma ^{(sph)}_{pol}(\theta )$ [Eq. (\ref{Eq.pol-charge-FULL})].
The most sensitive region is the one where $r\rightarrow r_{0}$.
There, for $r_{0}=8\sigma$ we have a relative error of $10^{-6}$ with $l_{max}=100$.
\bibitem{Note_Qr_resolution}
For the net fluid charge $Q(r) $ one has an arbitrary resolution
$\Delta r$.
\bibitem{Note_Linse_c}
This feature also shows how  important it is to treat precisely the self-image
interaction as done in the present study  with a Legendre sum.
\end{thebibliography}

\end{document}